\documentclass[preprint,prd,showpacs,showkeys]{revtex4}
\usepackage{amsmath,epsfig}
\usepackage{amsfonts}
\usepackage{color}
\usepackage{amssymb}
\usepackage{graphicx}
\usepackage[cmtip,arrow]{xy}
\usepackage{url}
\usepackage{enumerate}

\usepackage{amssymb,amsopn}

\numberwithin{equation}{section}

\def\la{\lambda}

\def\be{\beta}

\def\p{\partial}

\def\p{\partial}

\def\be{\begin{equation}}
\def\ee{\end{equation}}
\def\bea{\begin{eqnarray}}
\def\eea{\end{eqnarray}}

\newcounter{orange}
\renewcommand{\theorange}{\alph{orange}}
\begin{document}
\title{The  Weyl-Wigner-Moyal formalism on a discrete phase space. \\ I. A Wigner function for a nonrelativistic particle with spin}

\author{Maciej Przanowski}\email{maciej.przanowski@p.lodz.pl}
\affiliation{Professor Emeritus}

\author{Jaromir Tosiek}\email{jaromir.tosiek@p.lodz.pl}
\affiliation{Institute of Physics, {\L}\'{o}d\'{z} University of
Technology, W\'{o}lcza\'{n}ska 219, 90-924 {\L}\'od\'{z}, Poland.}

\author{Francisco J. Turrubiates}\email{fturrub@esfm.ipn.mx}
\affiliation{Departamento de F\'{\i}sica, Escuela Superior de F\'{\i}sica
y Matem\'aticas,\\ Instituto Polit\'ecnico Nacional, Unidad Adolfo
L\'opez Mateos, Edificio 9, 07738 Ciudad de M\'exico, M\'exico.}
\date{\today}

\begin{abstract}
The Weyl-Wigner-Moyal formalism for quantum particle with discrete internal degrees of freedom is developed.
A one to one correspondence between operators in the Hilbert space $L^{2}(\mathbb{R}^{3})\otimes{\mathcal{H}}^{(s+1)}$ and functions on the phase space $\mathbb{R}^{3}\times\mathbb{R}^{3}\times \{0,...,s\} \times\{0,...,s\}$ is found.
The expressions for the Stratonovich-Weyl quantizer, star product and Wigner functions of such systems for arbitrary
values of spin are obtained in detail. As examples the Landau levels and the corresponding Wigner functions for a spin $\frac{1}{2}$ nonrelativistic particle as well as the magnetic resonance for a spin $\frac{1}{2}$ nonrelativistic uncharged particle are analysed.
\end{abstract}

\keywords{Quantum mechanics in continuous and discrete phase space, Wigner function, spin.}

\vskip -1truecm
\pacs{03.65 Ca, 03.65 Aa, 03.65 Sq.}

\maketitle

\vskip -1.3truecm
\newpage

\setcounter{equation}{0}

\section{Introduction}

The phase space quantum description is an alternative approach to the usual construction of quantum mechanics in a
Hilbert space.  Its origins can be traced to the seminal
works of Weyl \cite{Weyl1,Weyl2}, Wigner \cite{Wigner}, Groenewold \cite{Groenewold}, Moyal \cite{Moyal}, and Vey \cite{Vey}. Physical systems in that formalism are modelled on  symplectic manifolds usually identified with  phase spaces of classical counterparts.

Information about the system is acquired by a real valued Wigner function which plays an analogous
role as the wave function.
For a complete review of this topic
see Refs. \cite{Kim,ZFC} and the references cited therein.
The Wigner function has found wide interest in various branches of physics such as nuclear and particle physics,
quantum optics, statistical physics and condensed matter physics \cite{Weinbub}. Moreover, several procedures have been proposed
for the experimental measurement of Wigner functions \cite{Kurtsiefer,Ourjoumtsev,Deleglise}.

Observables in phase space version of quantum mechanics are represented by real functions on the phase space. But on the contrary to classical physics multiplication of these functions is in general nonAbelian. Under this perspective transition from classical to quantum physics is
understood as a noncommutative deformation of the usual product algebra of the smooth functions on the classical phase space, which in
turn induces a modification of the Poisson bracket algebra.  Mathematical foundations of this attempt called \textit{deformation quantisation} were introduced by
Bayen et al in 1978 \cite{Bayen}.

The deformed multiplication is called the \textit{$*$  - product (star product)} and its existence was proved for the case of any symplectic manifold in  \cite{Wilde,Fedosov}. Even more, a method to obtain an explicit expression for such a $*$ - product was given by Fedosov by means of a geometrical construction \cite{Fedosov}.
Later, Kontsevich shown that the $*$ - product also exists for any Poisson manifold \cite{Kontsevich}. These results
in principle allow us to perform the quantisation of arbitrary Poissonian or symplectic systems which gives an advantage
over the other quantisation methods developed until now. An updated and detailed review of this quantisation procedure can be
consulted in Ref. \cite{DitoStern}.

Important results in the phase space approach have been obtained mainly for classical degrees
of freedom. Serious obstacles appear if one tries to deal with purely quantum ones. In particular,
the treatment of spin in this framework is a non-trivial problem.  A pioneering work on this problem was done by V\'arilly and Gracia-Bond\'ia. They considered the sphere ${S}^{2}$ as the corresponding phase space (see Ref. \cite{Varilly}).

We propose  a different path. In  the recent paper \cite{Przan1} we have investigated relationships between the Hilbert space version of quantum mechanics and its phase space representation  for finite
dimensional Hilbert spaces ${\mathcal{H}}^{(s+1)}$, $s=1,2,...$. Mathematical formulation of this relationship is called  the Weyl-Wigner formalism. Using widely the results
obtained previously by several authors (see Refs. from 1 to 40 in \cite{Przan1}) we were able to find
the general Weyl correspondence between operators in a finite dimensional Hilbert space
and functions on the respective discrete phase space as well as the natural definition of the Wigner function.
The results obtained are specified to the case of \textit{symmetric ordering} of operators
(when the dimension of Hilbert space is odd) and to the case of \textit{almost symmetric
ordering} (when the dimension of Hilbert space is even). It is evident that all these results
open the door to definition of the star product  and, consequently, to the
discrete Weyl-Wigner-Moyal formalism for finite dimensional Hilbert spaces. The phase
spaces associated to these Hilbert spaces are grids with finite number of points.

  The standard
``continuous" Weyl-Wigner-Moyal formalism for the Hilbert space $L^{2}(\mathbb{R}^{n})$ has been
an object of analysis for many years and now it is worked out with all details \cite{Weyl1,Weyl2,Wigner,Groenewold,Moyal,Bayen,Cohen,Agarwal,Tatarskij,Stratonovich,Fano,Hillery,Plebanski1,Plebanski2,Plebanski3,ZFC}.
Joining both ``discrete" and ``continuous" Weyl-Wigner-Moyal formalisms one gets the framework
for the tensor product of Hilbert spaces $L^{2}(\mathbb{R}^{n})\otimes{\mathcal{H}}^{(s+1)}$.
In consequence it is possible to develop quantum mechanics (nonrelativistic or relativistic) of
systems with internal degrees of freedom (eg. spin) within the phase space language.

Our present
and subsequent papers are devoted to this issue. Here we are going to focus on studying the
Weyl-Wigner-Moyal formalism on $L^{2}(\mathbb{R}^{3})\otimes{\mathcal{H}}^{(s+1)}$, $s=1,2,...$.
Then as examples of applications of that formalism we consider first the Landau levels and the
respective Wigner functions for a spin $\frac{1}{2}$ nonrelativistic charged particle, and later the
magnetic resonance for the spin $\frac{1}{2}$ nonrelativistic uncharged particle.

  Landau levels in terms
of the Weyl-Wigner-Moyal formalism but for the spinless nonrelativistic particle were investigated in all
details by B. Demircio\v{g}lu and A. Ver\c{c}in in their nice work \cite{Demircioglu}. The analogous
considerations in noncommutative spaces have been done by \"{O}. F. Dayi and L. T. Kelleyane \cite{Dayi}.
An interesting phase space approach to the spin $\frac{1}{2}$ magnetic dipole in an external magnetic
field and to the magnetic resonance has been given by P. Watson and A. J. Bracken \cite{Watson}. Semiclassical approach to
the quantum dynamics of a $\frac{1}{2}$ spin particle with the use of a Wigner-Weyl-type calculus under the assumption that
the phase space is $\mathbb{R}^{d}\times \mathbb{R}^{d}\times S^{2}$ has been discussed by O. Gat, M. Lein and S. Teufel
in an interesting paper \cite{Lein}. In this work the phase space picture of the Stern-Gerlach experiment is also proposed.

It is important to point out that quantisation in discrete phase spaces could be useful not only for spin treatment but also in other physical problems where finite-dimensional Hilbert spaces are involved. A promising example is the so-called polymer quantisation used in loop quantum gravity. When applying this quantisation technique to mechanical systems with a finite number of degrees of freedom, the result is not equivalent to the standard Schr\"{o}dinger quantisation and it is known as polymer representation. This description is fundamental in certain cosmological models and allows us to test some of the elements of loop quantum gravity in a simple way. This type of quantisation is currently an important topic of research and a detailed review of its construction as well as some valuable results can be consulted in \cite{Ashtekar1,Ashtekar2,Husain,Kunstatter}.

Our paper is organized as follows. In Section 2 we study the general Weyl correspondence between
operators in the Hilbert space $L^{2}(\mathbb{R}^{3})\otimes{\mathcal{H}}^{(s+1)}$, $s=1,2,...$ and
 functions on the respective phase space $\mathbb{R}^{3}\times\mathbb{R}^{3}\times\Gamma^{(s+1)}$
(see (\ref{216})). To this end we employ the unitary operators $\widehat{\mathcal{U}}(\lambda,\mu)$, $\lambda,\mu\in\mathbb{R}^{3}$,
on $L^{2}(\mathbb{R}^{3})$, and the Schwinger unitary operators $\widehat{\mathcal{D}}(k,l)$, $0\leq k,l\leq s$, on
${\mathcal{H}}^{(s+1)}$. These operators enable us to define the phase-point operators called the \textit{Stratonovich-Weyl
quantizer} or the \textit{Fano operators}, which next are used to define the Weyl correspondence.
In Sec. 3 the expression of the $*$-product on $\mathbb{R}^{3}\times\mathbb{R}^{3}\times\Gamma^{(s+1)}$ is obtained. Section 4 is
devoted to a natural determination of Wigner function. Then the properties of this function
are analysed and the evolution equation i.e. the Liouville-von Neumann-Wigner equation is given. In Section 5 we
study various possibilities of choosing the kernels $(\mathcal{P},\mathcal{K})$ determining the Stratonovich-Weyl
quantizer. Finally, we decide to choose $\mathcal{P}=1$ and $\mathcal{K}$ of the form (\ref{54}). Then we get the Wigner function in
the form given by (\ref{57}) or (\ref{59}) which is the main result of our paper.

In Sec. 6 we consider two examples of application of our formalism. Example A is devoted to the phase space analysis of
the spin $\frac{1}{2}$ nonrelativistic charged particle in a homogeneous magnetic field. Using the Weyl-Wigner-Moyal
formalism we find the Landau levels and corresponding Wigner functions. In  example B we study the magnetic
resonance for the spin $\frac{1}{2}$ nonrelativistic uncharged particle in terms of the Weyl-Wigner-Moyal formalism.
The Rabi frequency, resonance frequency and the Wigner functions are found. Finally, a brief summary (Sec. 7)
and the list of references close the paper.


\section{Preliminary considerations}

Motivated by the Pauli equation and by the relativistic Dirac equation we consider a Hilbert space of the form
\begin{equation}\label{21}
  \mathcal{H}=L^{2}(\mathbb{R}^{3})\otimes{\mathcal{H}}^{(s+1)}, \quad s=0,1,...
\end{equation}
where ${\mathcal{H}}^{(s+1)}$ is an $(s+1)$-dimensional Hilbert space.

Let $\{|n\rangle\}^{s}_{n=0}$ be some orthonormal basis of ${\mathcal{H}}^{(s+1)}$ i.e., $\langle n|n'\rangle=\delta_{nn'}$, $n,n'=0,...,s$.
Following the works on discrete Weyl-Wigner-Moyal formalism (see \cite{Przan1} and the extensive list of references therein) other orthonormal
basis $\{|\phi_{m}\rangle\}^{s}_{m=0}$ of ${\mathcal{H}}^{(s+1)}$ is given by
\begin{equation}\label{22}
  |\phi_{m}\rangle\:=\frac{1}{\sqrt{s+1}}\sum_{n=0}^{s}\exp\{i n \phi_{m}\} |n\rangle, \,\,\, m=0,...,s
\end{equation}
with
\begin{equation}\label{23}
  \phi_{m}=\phi_{0}+\frac{2\pi}{s+1}m, \,\,\, m=0,...,s.
\end{equation}
Without any loss of generality one can put
\begin{equation}\label{24}
  \phi_{0}=0
\end{equation}
and further on we assume that (\ref{24}) holds true.

Then we define two Hermitian operators
\begin{equation}\label{25}
  \widehat{n}:=\sum_{n=0}^{s}n|n\rangle\langle n|\,\,\, , \,\,\, \widehat{\phi}:=\sum_{m=0}^{s}\phi_{m}|\phi_{m}\rangle\langle \phi_{m}|
\end{equation}
which enable us to define two unitary generators, the \textit{Schwinger operators} (see \cite{Przan1} and the references therein)
\begin{equation}\label{26}
  \widehat{V}:=\exp\left\{i \frac{2\pi}{s+1}\widehat{n}\right\}, \,\,\, \widehat{R}:=\exp\{i\widehat{\phi}\}.
\end{equation}
One quickly finds the relation
\begin{equation}\label{27}
  \widehat{V}^{s+1}=\widehat{R}^{s+1}=\widehat{1}
\end{equation}
and the commutation rule
\begin{equation}\label{28}
  \exp\left\{-i\frac{\pi k l}{s+1}\right\}\widehat{R}^{k}\widehat{V}^{l}=\exp\left\{i\frac{\pi k l}{s+1}\right\}\widehat{V}^{l}\widehat{R}^{k}
\end{equation}
for any $k,l \in \mathbb{Z}$.
We introduce the unitary operators
\begin{equation}\label{29}
  \widehat{\mathcal{D}}(k,l):=\exp\left\{-i\frac{\pi k l}{s+1}\right\}\widehat{R}^{k}\widehat{V}^{l}, \,\,\, k,l \in \mathbb{Z}.
\end{equation}
These operators will play a crucial role in further considerations. One can easily show that they fulfill the following relations
\begin{subequations}\label{210}
\begin{align}
   \widehat{\mathcal{D}}^{\dag}(k,l)&= \widehat{\mathcal{D}}^{-1}(k,l)=\widehat{\mathcal{D}}(-k,-l),  \,\,\, k,l \in \mathbb{Z} \label{210a}\\
   \textrm{Tr}\left\{\widehat{\mathcal{D}}(k,l) \right\} &= (s+1)\delta_{k0}\delta_{l0}, \,\,\, 0\leq k,l\leq s \label{210b}\\
   \textrm{Tr}\left\{\widehat{\mathcal{D}}(k,l)\widehat{\mathcal{D}}^{\dag}(k',l') \right\} &= (s+1)\delta_{kk'}\delta_{ll'}, \,\,\, 0\leq k,l,k',l'\leq s. \label{210c}
\end{align}
\end{subequations}
It is convenient to extend the definition (\ref{22}) to all $m \in \mathbb{Z}$ by putting (remember about (\ref{24}))
\begin{equation}\label{211}
  |\phi_{m}\rangle:=\frac{1}{\sqrt{s+1}}\sum_{n=0}^{s}\exp\left\{i\frac{2\pi}{s+1}nm\right\}|n\rangle, \,\,\, m \in \mathbb{Z},
\end{equation}
\[
\langle \phi_{p}|\phi_{m}\rangle= \delta_{p\,m \,{\rm mod}(s+1)}.
\]
Employing (\ref{211}) we quickly find that the operators $\widehat{\mathcal{D}}(k,l)$ for $0\leq k,l\leq s$ can be represented in the form
\begin{subequations}\label{212}
\begin{align}
  \widehat{\mathcal{D}}(k,l)=\exp\left\{i\frac{\pi k l}{s+1}\right\}\sum_{m=0}^{s}\exp\left\{i\frac{2\pi k m}{s+1}\right\}|\phi_{m+l}\rangle\langle\phi_{m}|, \\
  \label{212a}
 \widehat{\mathcal{D}}(k,l)=\exp\left\{i\frac{\pi k l}{s+1}\right\}
   \left( \sum_{n=0}^{s-k} \exp\left\{i\frac{2\pi n l}{s+1}\right\}
   |n \rangle \langle k+n|+ \right.
   \nonumber \\
   \left.
     \sum_{n=s-k+1}^{s}
   \exp\left\{i\frac{2\pi n l}{s+1}\right\}
   |n \rangle \langle k+n-s-1|
   \right).
   \end{align}
  \end{subequations}
In the next step one defines the unitary operators on $L^{2}(\mathbb{R}^{3})$ (the \textit{displacement operators}) as (see eg. \cite{Weyl2,Tatarskij,Plebanski2,Plebanski3,ZFC})
\begin{eqnarray}\label{213}
  \widehat{\mathcal{U}}(\lambda,\mu):&=&\exp\{i(\lambda\cdot\widehat{p}+\mu\cdot\widehat{q})\} \nonumber \\
   &=& \exp\left\{-i \frac{\hbar \lambda\cdot\mu}{2}\right\}\exp\{i\lambda\cdot\widehat{p}\}\exp\{i\mu\cdot\widehat{q}\} \nonumber \\
   &=& \exp\left\{i \frac{\hbar \lambda\cdot\mu}{2}\right\}\exp\{i\mu\cdot\widehat{q}\}\exp\{i\lambda\cdot\widehat{p}\}
\end{eqnarray}
where, here and in all the paper the following abbreviations are used:
$\lambda= (\lambda_1, \lambda_2, \lambda_3) \in {\mathbb R}^3$,
$\mu= (\mu_1, \mu_2, \mu_3) \in {\mathbb R}^3$,
 $\widehat{q}=(\widehat{q}_{1},\widehat{q}_{2},\widehat{q}_{3})$,
$\widehat{p}=(\widehat{p}_{1},\widehat{p}_{2},\widehat{p}_{3})$, $\lambda\cdot\widehat{p}=\lambda_{1}\widehat{p}_{1}+\lambda_{2}\widehat{p}_{2}+\lambda_{3}\widehat{p}_{3}$,
$\mu\cdot\widehat{q}=\mu_{1}\widehat{q}_{1}+\mu_{2}\widehat{q}_{2}+\mu_{3}\widehat{q}_{3}$, $\lambda\cdot\mu=\lambda_{1}\mu_{1}+\lambda_{2}\mu_{2}+\lambda_{3}\mu_{3}$ etc.
The operators $\widehat{\mathcal{U}}(\lambda,\mu)$ have the analogous properties as the operators $\widehat{\mathcal{D}}(k,l)$ (see (\ref{210}))
\begin{subequations}\label{214}
\begin{align}
   \widehat{\mathcal{U}}^{\dag}(\lambda,\mu)&= \widehat{\mathcal{U}}^{-1}(\lambda,\mu)=\widehat{\mathcal{U}}(-\lambda,-\mu), \label{214a}\\
   \textrm{Tr}\left\{\widehat{\mathcal{U}}(\lambda,\mu) \right\} &= \left(\frac{2\pi}{\hbar}\right)^{3}\delta(\lambda)\delta(\mu), \label{214b}\\
   \textrm{Tr}\left\{\widehat{\mathcal{U}}(\lambda,\mu)\widehat{\mathcal{U}}^{\dag}(\lambda',\mu') \right\} &= \left(\frac{2\pi}{\hbar}\right)^{3}\delta(\lambda-\lambda')\delta(\mu-\mu') \label{214c}
\end{align}
\end{subequations}
and they can be written in the form
\begin{eqnarray}\label{215}
  \widehat{\mathcal{U}}(\lambda,\mu)&=& \int\limits_{\mathbb{R}^{3}}\exp\{i\mu\cdot q\}|q-\frac{\hbar\lambda}{2}\rangle dq \langle q+\frac{\hbar\lambda}{2}|, \nonumber \\
      &=& \int\limits_{\mathbb{R}^{3}}\exp\{i\lambda\cdot p\}|p+\frac{\hbar\mu}{2}\rangle dp \langle p-\frac{\hbar\mu}{2}|
\end{eqnarray}
with $dq:=dq_1 dq_2 dq_3\, , \, dp:=dp_1 dp_2 dp_3.$

As it is done in the case of ``\textit{usual continuous}" Weyl-Wigner-Moyal formalism we should find now the phase space associated to the Hilbert
space $\mathcal{H}$ given by (\ref{21}). Using the results of previous works (\cite{Przan1} and the references therein; \cite{Weyl1,Weyl2,Wigner,Groenewold,Moyal,Bayen,Cohen,Agarwal,Tatarskij,Stratonovich,Fano,Hillery,Plebanski1,Plebanski2,Plebanski3,ZFC}) it is obvious that the natural
phase space is here
\begin{equation}\label{216}
  \Gamma:=\{(p,q,\phi_{m},n)\}=\mathbb{R}^{3}\times\mathbb{R}^{3}\times\Gamma^{(s+1)}
\end{equation}
where $\Gamma^{(s+1)}$ is the $(s+1)\times(s+1)$ grid, $\Gamma^{(s+1)}=\{(\phi_{m},n)\}_{m,n=0,...,s}$.

Let $\widehat{f}$ be a linear operator acting in $\mathcal{H}$. Employing (\ref{212}) and (\ref{215}) one easily gets
$$
\textrm{Tr}\left\{\widehat{f}\widehat{\mathcal{U}}^{\dag}(\lambda,\mu)\widehat{\mathcal{D}}^{\dag}(k,l)\right\} = \exp\left\{-i \left(\frac{\hbar\lambda\cdot\mu}{2}+\frac{\pi k l}{s+1} \right)\right\}\sum_{m,n=0}^{s}\int_{\mathbb{R}^{3}\times\mathbb{R}^{3}}dpdq \exp\{-i (\lambda\cdot p + \mu\cdot q)\}
$$
\begin{equation}
 \exp\left\{-i \frac{2\pi}{s+1}(km+ln) \right\}\langle p,\phi_{m}|q,n\rangle \langle q,n |\widehat{f}|p,\phi_{m}\rangle \nonumber
\end{equation}
$$
=\exp\left\{i \left(\frac{\hbar\lambda\cdot\mu}{2}+\frac{\pi k l}{s+1} \right)\right\}\sum_{m,n=0}^{s}\int_{\mathbb{R}^{3}\times\mathbb{R}^{3}}dpdq \exp\{-i (\lambda\cdot p + \mu\cdot q)\}
$$
\begin{equation}\label{217}
\exp\left\{-i \frac{2\pi}{s+1}(km+ln) \right\}\langle q,n |p,\phi_{m}\rangle \langle p,\phi_{m}|\widehat{f}|q,n\rangle
\end{equation}
for any $\lambda \in \mathbb{R}^{3}$, $\mu \in \mathbb{R}^{3}$ and $0\leq k,l\leq s$; where $\widehat{\mathcal{U}}^{\dag}(\lambda,\mu)\widehat{\mathcal{D}}^{\dag}(k,l):=\widehat{\mathcal{U}}^{\dag}(\lambda,\mu)\otimes\widehat{\mathcal{D}}^{\dag}(k,l)$, $|p,\phi_{m}\rangle:=|p\rangle \otimes |\phi_{m}\rangle$, $|q,n\rangle:=|q\rangle \otimes |n\rangle$, etc.

In order to establish coefficient convention we give the Fourier transformation formulas  applied in our paper:
\setcounter{orange}{1}
\renewcommand{\theequation} {\arabic{section}.\arabic{equation}\theorange}
\begin{enumerate}
\item
the Fourier transform
\be \label{dod1}
\tilde{F}(\lambda, \mu):= \int\limits_{\mathbb{R}^{3}\times\mathbb{R}^{3}}dpdq \exp\{-i (\lambda\cdot p + \mu\cdot q)\}F(p,q),
\ee
\addtocounter{orange}{1}
\addtocounter{equation}{-1}
\item
the inverse Fourier transform
\be \label{dod2}
F(p,q):= \frac{1}{(2 \pi)^6}
\int\limits_{\mathbb{R}^{3}\times\mathbb{R}^{3}}d \lambda d \mu \exp\{i (\lambda\cdot p + \mu\cdot q)\}
\tilde{F}(\lambda, \mu),
\ee
\addtocounter{orange}{1}
\addtocounter{equation}{-1}
\item
the discrete Fourier transform
\be \label{dod3}
\tilde{f}(k,l):= \sum_{m,n=0}^s
\exp\left\{-i \frac{2\pi}{s+1}(km+ln) \right\} f(m,n)
\ee
\addtocounter{orange}{1}
\addtocounter{equation}{-1}
\item
the inverse discrete Fourier transform
\be \label{dod4}
f(m,n):= \frac{1}{(s+1)^2} \sum_{k,l=0}^s
\exp\left\{i \frac{2\pi}{s+1}(km+ln) \right\}\tilde{f}(k,l).
\ee
\end{enumerate}
\renewcommand{\theequation} {\arabic{section}.\arabic{equation}}

It is quite clear that Eq. (\ref{217}) gives the Fourier transformations of the functions: $\langle p,\phi_{m}|q,n\rangle \langle q,n |\widehat{f}|p,\phi_{m}\rangle$ and $\langle q,n |p,\phi_{m}\rangle \langle p,\phi_{m}|\widehat{f}|q,n\rangle$, which are defined on the phase space $\Gamma$. So, the inverse transformations to (\ref{217}) give
\begin{subequations}\label{218}
\begin{align}
\begin{split}
\langle p,\phi_{m}|q,n\rangle \langle q,n |\widehat{f}|p,\phi_{m}\rangle =& \frac{1}{(2\pi)^{6}(s+1)^{2}}\sum_{k,l=0}^{s}\int_{\mathbb{R}^{3}\times\mathbb{R}^{3}}d\lambda d\mu \exp\left\{i\frac{\hbar\lambda\cdot\mu}{2}\right\}\exp\left\{i(\lambda\cdot p + \mu\cdot q)\right\} \\
&\exp\left\{i \frac{\pi k l}{s+1}\right\}\exp\left\{i\frac{2\pi}{s+1}(km+ln)\right\} \textrm{Tr}\left\{\widehat{f}\widehat{\mathcal{U}}^{\dag}(\lambda,\mu)\widehat{\mathcal{D}}^{\dag}(k,l)\right\}, \label{218a}
\end{split}
\\
\begin{split}
\langle q,n |p,\phi_{m}\rangle \langle p,\phi_{m}|\widehat{f}|q,n\rangle =& \frac{1}{(2\pi)^{6}(s+1)^{2}}\sum_{k,l=0}^{s}\int_{\mathbb{R}^{3}\times\mathbb{R}^{3}}d\lambda d\mu \exp\left\{-i\frac{\hbar\lambda\cdot\mu}{2}\right\}\exp\left\{i(\lambda\cdot p + \mu\cdot q)\right\} \\
&\exp\left\{-i \frac{\pi k l}{s+1}\right\}\exp\left\{i\frac{2\pi}{s+1}(km+ln)\right\} \textrm{Tr}\left\{\widehat{f}\widehat{\mathcal{U}}^{\dag}(\lambda,\mu)\widehat{\mathcal{D}}^{\dag}(k,l)\right\}, \label{218b}
\end{split}
\end{align}
\end{subequations}
where $d \lambda:=d \lambda_1 d \lambda_2 d \lambda_3$ and $d \mu:= d\mu_1 d\mu_2 d \mu_3.$

Now the following correspondences between functions on $\Gamma$ and linear operators in $\mathcal{H}$ can be defined
$$
  f(p,q,\phi_{m},n) \,\, \longleftrightarrow \widehat{f}
$$
\begin{subequations}\label{219}
\begin{align}
  f(p,q,\phi_{m},n)=(2\pi\hbar)^{3}(s+1)\langle p,\phi_{m}|q,n\rangle \langle q,n |\widehat{f}|p,\phi_{m}\rangle \label{219a} \\
\intertext{or}
  f(p,q,\phi_{m},n)=(2\pi\hbar)^{3}(s+1)\langle q,n |p,\phi_{m}\rangle \langle p,\phi_{m}|\widehat{f}|q,n\rangle \label{219b}
\end{align}
\end{subequations}
where the factor $(2\pi\hbar)^{3}(s+1)$ is used to ensure the relation $1 \leftrightarrow \widehat{1}$. Both correspondences, (\ref{219a}) and (\ref{219b}), are one to one.

We can easily note that (\ref{219a}) leads to the \textit{standard ordering of operators} i.e. the operators $\widehat{q}$ and $\widehat{n}$ are to the left of $\widehat{p}$ and $\widehat{\phi}$. Analogously (\ref{219b}) gives the \textit{anti-standard ordering of operators} what means that $\widehat{p}$ and $\widehat{\phi}$ are to the left of $\widehat{q}$ and $\widehat{n}$.

The relations (\ref{219}) are direct and obvious generalizations of the results of P.A.M. Dirac \cite{Dirac} for the continuous case and of S. Chaturvedi et al \cite{Chaturvedi} for the discrete case on the case of Hilbert space $\mathcal{H}$ given by (\ref{21}).
From (\ref{217}), (\ref{219a}) and (\ref{219b}) with (\ref{dod1}), (\ref{dod3}) we conclude that the Fourier transform $\tilde{f}(\lambda,\mu,k,l)$ of $f(p,q,\phi_{m},n)$ equal to
\begin{equation}\label{220}
  \tilde{f}(\lambda,\mu,k,l):=\sum_{m,n=0}^{s}\int_{\mathbb{R}^{3}\times\mathbb{R}^{3}}dpdq f(p,q,\phi_{m},n) \exp\{-i(\lambda\cdot p +\mu\cdot q)\}\exp\left\{ -i\frac{2\pi}{s+1}(km+ln)\right\}
\end{equation}
reads
\begin{subequations}\label{221}
\begin{align}
  \tilde{f}(\lambda,\mu,k,l)=(2\pi\hbar)^{3}(s+1)\exp\left\{i\left(\frac{\hbar \lambda\cdot\mu}{2}+\frac{\pi k l}{s+1}\right)\right\} \textrm{Tr}\left\{\widehat{f}\widehat{\mathcal{U}}^{\dag}(\lambda,\mu)\widehat{\mathcal{D}}^{\dag}(k,l)\right\},& \label{221a} \\
 \lambda,\mu \in \mathbb{R}^{3}, 0\leq k,l\leq s \;\;\;\;\;\;\;\;\;\;\;\;\;\;\;\;\;\;\;\;\;\;\;\;\;\;\;\;\;\;\;\;& \nonumber
\intertext{for $f$ given by (\ref{219a}), and}
  \tilde{f}(\lambda,\mu,k,l)=(2\pi\hbar)^{3}(s+1)\exp\left\{-i\left(\frac{\hbar \lambda\cdot\mu}{2}+\frac{\pi k l}{s+1}\right)\right\} \textrm{Tr}\left\{\widehat{f}\widehat{\mathcal{U}}^{\dag}(\lambda,\mu)\widehat{\mathcal{D}}^{\dag}(k,l)\right\},  &\label{221b} \\
  \lambda,\mu \in \mathbb{R}^{3}, 0\leq k,l\leq s \;\;\;\;\;\;\;\;\;\;\;\;\;\;\;\;\;\;\;\;\;\;\;\;\;\;\;\;\;\;\;\; & \nonumber
\end{align}
\end{subequations}
for $f$ as in (\ref{219b}).
\\
A quick glance at Eqs. (\ref{221}) shows that one can define a one to one correspondence between functions (distributions) on $\mathbb{R}^{3}\times\mathbb{R}^{3}\times \{0,...,s\} \times\{0,...,s\}$ and linear operators in $\mathcal{H}$ as follows
$$
  \tilde{\tilde{f}}=\tilde{\tilde{f}}(\lambda,\mu,k,l) \longleftrightarrow \widehat{f}, \,\,\, \lambda,\mu \in \mathbb{R}^{3}; k,l=0,...,s,
$$
$$
\widehat{f} = \sum_{k,l=0}^{s}\int_{\mathbb{R}^{3}\times\mathbb{R}^{3}}d\lambda d\mu  \tilde{\tilde{f}}(\lambda,\mu,k,l) \widehat{\mathcal{U}}(\lambda,\mu)\widehat{\mathcal{D}}(k,l),
$$
\begin{equation}\label{222}
\tilde{\tilde{f}}(\lambda,\mu,k,l) = \left(\frac{\hbar}{2\pi}\right)^{3}(s+1)^{-1} \textrm{Tr}\left\{\widehat{f}\widehat{\mathcal{U}}^{\dag}(\lambda,\mu)\widehat{\mathcal{D}}^{\dag}(k,l)\right\}.
\end{equation}
The factor $\left(\frac{\hbar}{2\pi}\right)^{3}(s+1)^{-1}$ is taken so that
\begin{equation}\label{223}
  \delta(\lambda)\delta(\mu)\delta_{k0}\delta_{l0} \longleftrightarrow \widehat{1}
\end{equation}
(see (\ref{210b}) and (\ref{214b})).
\\
Let $\tilde{\tilde{f}}= \tilde{\tilde{f}}(\lambda,\mu,k,l)$ and $\tilde{\tilde{g}}= \tilde{\tilde{g}}(\lambda,\mu,k,l)$ be the functions associated to linear operators $\widehat{f}$ and $\widehat{g}$, respectively, according to the recipe (\ref{222}).
We are searching for the function which corresponds to the product of operators $\widehat{f}\cdot\widehat{g}$. Denote this function by $(\tilde{\tilde{f}}\boxtimes\tilde{\tilde{g}})(\lambda,\mu,k,l)$.
\\
From (\ref{222}) one has
$$
(\tilde{\tilde{f}}\boxtimes\tilde{\tilde{g}})(\lambda,\mu,k,l)= \left(\frac{\hbar}{2\pi}\right)^{3}(s+1)^{-1} \textrm{Tr}\left\{\widehat{f}\cdot\widehat{g}\;\;\widehat{\mathcal{U}}^{\dag}(\lambda,\mu)\widehat{\mathcal{D}}^{\dag}(k,l)\right\}
$$
$$
= \left(\frac{\hbar}{2\pi}\right)^{3}(s+1)^{-1}\sum_{k',l',k'',l''=0}^s \int_{\mathbb{R}^{3}\times\mathbb{R}^{3}\times\mathbb{R}^{3}\times\mathbb{R}^{3}}d\lambda'd\mu'd\lambda''d\mu''\tilde{\tilde{f}}(\lambda',\mu',k',l')
$$
\begin{equation}\label{224}
 \textrm{Tr}\left\{\widehat{\mathcal{U}}(\lambda',\mu')\widehat{\mathcal{D}}(k',l')\cdot\widehat{\mathcal{U}}(\lambda'',\mu'')\widehat{\mathcal{D}}(k'',l'')
\cdot\widehat{\mathcal{U}}^{\dag}(\lambda,\mu)\widehat{\mathcal{D}}^{\dag}(k,l)\right\}\tilde{\tilde{g}}(\lambda'',\mu'',k'',l'').
\end{equation}
Straightforward calculations based on relations (\ref{212}) and (\ref{215}) give
$$
\textrm{Tr}\left\{\widehat{\mathcal{U}}(\lambda',\mu')\widehat{\mathcal{D}}(k',l')\cdot\widehat{\mathcal{U}}(\lambda'',\mu'')\widehat{\mathcal{D}}(k'',l'')
\cdot\widehat{\mathcal{U}}^{\dag}(\lambda,\mu)\widehat{\mathcal{D}}^{\dag}(k,l)\right\}=\left(\frac{2\pi}{\hbar}\right)^{3}(s+1)
$$
$$
(-1)^{(k'-k)\theta(l'-l-1)+(l'-l)\theta(k'-k-1)+(s+1)\theta(k'-k-1)\theta(l'-l-1)}\exp\left\{\frac{i\hbar}{2}(\lambda'\cdot\mu-\lambda\cdot\mu')\right\}
$$
\begin{equation}\label{225}
\exp\left\{i\frac{\pi}{s+1}(k'l-kl')\right\}\delta(\lambda'+\lambda''-\lambda)\delta(\mu'+\mu''-\mu)\delta_{k'+k''-k,0mod(s+1)}\delta_{l'+l''-l,0mod(s+1)}
\end{equation}
where $\theta(j)$, $j\in\mathbb{Z}$, is the \textit{discrete Heaviside step function}
\begin{equation}\label{226}
  \theta(j)=\left\{ \begin{array}{rl}
 1, &\mbox{$j\geq0$} \\
 0, &\mbox{$j<0$}
       \end{array} \right. , \,\,\, j\in\mathbb{Z}
\end{equation}
Now we should note a grave lack in the function-operator correspondence given by (\ref{219}). Namely, the operator associated to a real function is, in general, non-Hermitian. In consequence neither (\ref{219a}) nor (\ref{219b}) can define the quantisation rule for a classical system. In order to improve this disadvantage we first rewrite the formulae (\ref{221}) as
\begin{equation}\label{227}
  \exp\left\{\mp i\frac{\hbar \lambda\cdot\mu}{2}\right\} \exp\left\{\mp i\frac{\pi k l}{s+1}\right\}\tilde{f}(\lambda,\mu,k,l)=(2\pi\hbar)^{3}(s+1)
\textrm{Tr}\left\{\widehat{f}\widehat{\mathcal{U}}^{\dag}(\lambda,\mu)\widehat{\mathcal{D}}^{\dag}(k,l)\right\}
\end{equation}
where the sign $``-"$ in exponents corresponds to (\ref{221a}) and $``+"$ to (\ref{221b}).

Perhaps the simplest direct generalization of (\ref{227}) reads
\begin{equation}\label{228}
  \mathcal{P}\left(\frac{\hbar \lambda\cdot\mu}{2}\right)\mathcal{K}\left(\frac{\pi k l}{s+1}\right)\tilde{f}(\lambda,\mu,k,l)=(2\pi\hbar)^{3}(s+1)
\textrm{Tr}\left\{\widehat{f}\widehat{\mathcal{U}}^{\dag}(\lambda,\mu)\widehat{\mathcal{D}}^{\dag}(k,l)\right\}
\end{equation}
where
\begin{equation}\label{229}
  \mathcal{K}\left(\frac{\pi k l}{s+1}\right)\neq 0  \,\,\,  \forall k,l \in \{0,...,s\}
\end{equation}
and
\begin{equation}\label{230}
  \mathcal{P}\left(\frac{\hbar \lambda\cdot\mu}{2}\right)\neq 0 \,\,\, \text{almost everywhere for } \lambda,\mu \in \mathbb{R}^{3}
\end{equation}
Thus we assume a new correspondence between the functions on $\Gamma$ and the operators in $\mathcal{H}$ (see (\ref{220}), (\ref{222}) and (\ref{228}))
$$
f(p,q,\phi_{m},n) \mathrel{\mathop{\longleftrightarrow}^{(\mathcal{P},\mathcal{K})}} \widehat{f}
$$
\begin{subequations}\label{231}
\begin{align}
\begin{split}
f(p,q,\phi_{m},n)=&\left(\frac{\hbar}{2\pi}\right)^{3}(s+1)^{-1}\sum_{k,l=0}^{s}\int_{\mathbb{R}^{3}\times\mathbb{R}^{3}}d\lambda d\mu\left(\mathcal{P}\left(\frac{\hbar \lambda\cdot\mu}{2}\right)\mathcal{K}\left(\frac{\pi k l}{s+1}\right)\right)^{-1}\\
&\exp\{i(\lambda\cdot p +\mu\cdot q)\}\exp\left\{ i\frac{2\pi}{s+1}(km+ln)\right\}
\textrm{Tr}\left\{\widehat{f}\widehat{\mathcal{U}}^{\dag}(\lambda,\mu)\widehat{\mathcal{D}}^{\dag}(k,l)\right\} \label{231a}
\end{split}
\\
\begin{split}
\widehat{f}=&\frac{1}{(2\pi)^{6}(s+1)^{2}}\sum_{k,l,m,n=0}^{s}\int_{\mathbb{R}^{3}\times\mathbb{R}^{3}\times\mathbb{R}^{3}\times\mathbb{R}^{3}}d\lambda d\mu dp dq \mathcal{P}\left(\frac{\hbar \lambda\cdot\mu}{2}\right)\mathcal{K}\left(\frac{\pi k l}{s+1}\right)\\
&\exp\{-i(\lambda\cdot p +\mu\cdot q)\}\exp\left\{-i\frac{2\pi}{s+1}(km+ln)\right\}f(p,q,\phi_{m},n)\widehat{\mathcal{U}}(\lambda,\mu)\widehat{\mathcal{D}}(k,l)\\
=& \frac{1}{(2\pi)^{3}(s+1)}\sum_{m,n=0}^{s}\int_{\mathbb{R}^{3}\times\mathbb{R}^{3}} dp dq f(p,q,\phi_{m},n)\widehat{\Omega}[\mathcal{P},\mathcal{K}](p,q,\phi_{m},n)\label{231b}
\end{split}
\end{align}
\end{subequations}
where
$$
\widehat{\Omega}[\mathcal{P},\mathcal{K}](p,q,\phi_{m},n):=\left(\frac{\hbar}{2\pi}\right)^{3}(s+1)^{-1}\sum_{k,l=0}^{s}\int_{\mathbb{R}^{3}\times\mathbb{R}^{3}}d\lambda d\mu
\mathcal{P}\left(\frac{\hbar \lambda\cdot\mu}{2}\right)\mathcal{K}\left(\frac{\pi k l}{s+1}\right)
$$
\begin{equation}\label{232}
\exp\{-i(\lambda\cdot p +\mu\cdot q)\}\exp\left\{-i\frac{2\pi}{s+1}(km+ln)\right\}
\widehat{\mathcal{U}}(\lambda,\mu)\widehat{\mathcal{D}}(k,l)
\end{equation}
defines the phase-point operators called the \textit{Stratonovich-Weyl quantizer} \cite{Stratonovich} or the \textit{Fano operators} \cite{Fano}.
\\
We impose the following natural conditions on the correspondence given by (\ref{231}):
\begin{enumerate}[(i)]
\item For any function $f$ on $\Gamma$ which depends on one variable i.e. $f=f(p)$, $f=f(q)$, $f=f(\phi_{m})$ or $f=f(n)$ the associated operator $\widehat{f}$ is exactly the same as in (\ref{219}) i.e. $\widehat{f}=f(\widehat{p})$, $\widehat{f}=f(\widehat{q})$, $\widehat{f}=f(\widehat{\phi})$ or $\widehat{f}=f(\widehat{n})$, respectively. This condition is fulfilled iff
\begin{equation}\label{233}
  \mathcal{P}(0)=1 \,\,\, \text{and} \,\,\, \mathcal{K}(0)=1
\end{equation}
\item For any real function $f(p,q,\phi_{m},n)$ the corresponding operator $\widehat{f}$ is Hermitian. This condition is satisfied iff
\begin{subequations}\label{234}
\begin{align}
    \mathcal{P}^{*}=\mathcal{P} \;\;\;\;\;\;\;\;\;\;\;\;\;\;\;\;\;\;\;\;\;\;\;\;\;\;\;\;\;\;\;\;\;\;\;\;\;\;\;\;\;\;\;\;\;\;\;\;\;
     \label{234a} \\
\intertext{and}
 \mathcal{K}^{*}\left(\frac{\pi k l}{s+1}\right)= (-1)^{s+1-k-l}\mathcal{K}\left(\frac{\pi(s+1-k)(s+1-l)}{s+1}\right), \,\,\, 1\leq k,l\leq s, \nonumber \\
\mathcal{K}^{*}(0)= \mathcal{K}(0) \;\;\;\;\;\;\;\;\;\;\;\;\;\;\;\;\;\;\;\;\;\;\;\;\;\;\;\;\;\;\;\;\;\;\;\;\;\;\;\;\;\;\;\;\;\;\;\;\;\label{234b}
\end{align}
\end{subequations}
where the asterisk ``${}^*$" stands for the complex conjugation.
\end{enumerate}
Further on in the paper we assume that (\ref{233}) and (\ref{234}) hold true. These assumptions together with the properties of  $ \widehat{\mathcal{U}}(\lambda,\mu)$ and $\widehat{\mathcal{D}}(k,l)$  yield
\begin{subequations}\label{235}
\begin{gather}
 \widehat{\Omega}^{\dag}[\mathcal{P},\mathcal{K}] = \widehat{\Omega}[\mathcal{P},\mathcal{K}] \label{235a} \\
 \textrm{Tr}\left\{\widehat{\Omega}[\mathcal{P},\mathcal{K}]\right\} = 1 \label{235b} \\
\begin{split}
\textrm{Tr}\left\{\widehat{\Omega}[\mathcal{P},\mathcal{K}](p,q,\phi_{m},n)\widehat{\Omega}[\mathcal{P},\mathcal{K}](p',q',\phi_{m'},n')\right\} = \\
\left(\frac{\hbar}{2\pi}\right)^{3}(s+1)^{-1}\sum_{k,l=0}^{s}\int_{\mathbb{R}^{3}\times\mathbb{R}^{3}}d\lambda d\mu
\left|\mathcal{P}\left(\frac{\hbar \lambda\cdot\mu}{2}\right)\mathcal{K}\left(\frac{\pi k l}{s+1}\right)\right|^{2} \\
\exp\{i[\lambda\cdot(p-p') +\mu\cdot (q-q')]\}\exp\left\{i\frac{2\pi}{s+1}[k(m-m')+l(n-n')]\right\} \label{235c}
\end{split}
\end{gather}
\end{subequations}
From (\ref{235c}) one quickly infers that
\begin{equation}\label{236}
\textrm{Tr}\left\{\widehat{\Omega}[\mathcal{P},\mathcal{K}](p,q,\phi_{m},n)\widehat{\Omega}[\mathcal{P},\mathcal{K}](p',q',\phi_{m'},n')\right\}=(2\pi\hbar)^{3}(s+1)
\delta(p-p')\delta(q-q')\delta_{mm'}\delta_{nn'}
\end{equation}
if and only if
\begin{eqnarray}\label{237}
  \left|\mathcal{P}\left(\frac{\hbar \lambda\cdot\mu}{2}\right) \right| &=& 1 \text{ almost everywhere on } \mathbb{R}^{3}\times\mathbb{R}^{3}; \nonumber \\
  \left|\mathcal{K}\left(\frac{\pi k l}{s+1}\right)\right|&=& 1 \,\,\, \forall \,\,\, 0\leq k,l\leq s
\end{eqnarray}
This last fact shows that it seems to be natural to assume that the conditions (\ref{237}) are fulfilled, and it is indeed usually assumed in the works on the subject.

However, since the choice of kernels $\mathcal{P}$ and $\mathcal{K}$ determines the ordering of operators and, as has been shown in our previous paper \cite{Przan1}, some
interesting results can be found under the choice of $\mathcal{K}$ such that the second requirement (\ref{237}) is not satisfied, we decide at the moment not to assume that it holds.

From (\ref{231b}) and (\ref{235c}) one gets
$$
f(p,q,\phi_{m},n)=\frac{1}{(2\pi)^{6}(s+1)^{2}}\sum_{k,l,m',n'=0}^{s}\int_{\mathbb{R}^{3}\times\mathbb{R}^{3}\times\mathbb{R}^{3}\times\mathbb{R}^{3}}d\lambda d\mu dp' dq'
\left|\mathcal{P}\left(\frac{\hbar \lambda\cdot\mu}{2}\right)\mathcal{K}\left(\frac{\pi k l}{s+1}\right)\right|^{-2}
$$
$$
\exp\{i[\lambda\cdot(p-p')+\mu\cdot(q-q')]\}\exp\left\{i\frac{2\pi}{s+1}[k(m-m')+l(n-n')]\right\}
$$
\begin{equation}\label{238}
\textrm{Tr}\left\{\widehat{f}\widehat{\Omega}[\mathcal{P},\mathcal{K}](p',q',\phi_{m'},n')\right\}.
\end{equation}
Hence
\begin{equation}\label{239}
  f(p,q,\phi_{m},n)=\textrm{Tr}\left\{\widehat{f}\;\widehat{\Omega}[\mathcal{P},\mathcal{K}](p,q,\phi_{m},n)\right\}
\end{equation}
for every $\widehat{f}$ iff (\ref{237}) holds true.
\section{Star product}
Let $f=f(p,q,\phi_{m},n)$ and $g=g(p,q,\phi_{m},n)$ be two functions on the phase space $\Gamma$ which correspond to the operators $\widehat{f}$ and $\widehat{g}$, respectively, according to the rule (\ref{231a}) or, equivalently, (\ref{238}). We are looking now for the function which corresponds to the product of operators $\widehat{f}\cdot\widehat{g}$.  This function will be denoted by $f*g=(f*g)(p,q,\phi_{m},n)$.

Employing the relation (\ref{231a}) one easily gets
$$
(f*g)(p,q,\phi_{m},n)=\left(\frac{\hbar}{2\pi}\right)^{3}(s+1)^{-1}\sum_{k,l=0}^{s}\int_{\mathbb{R}^{3}\times\mathbb{R}^{3}}d\lambda d\mu\left(\mathcal{P}\left(\frac{\hbar \lambda\cdot\mu}{2}\right)\mathcal{K}\left(\frac{\pi k l}{s+1}\right)\right)^{-1}
$$
$$
\exp\{i(\lambda\cdot p +\mu\cdot q)\}\exp\left\{ i\frac{2\pi}{s+1}(km+ln)\right\}
\textrm{Tr}\left\{\widehat{f}\cdot\widehat{g}\;\;\widehat{\mathcal{U}}^{\dag}(\lambda,\mu)\widehat{\mathcal{D}}^{\dag}(k,l)\right\}
$$
$$
=\sum_{k,l=0}^{s}\int_{\mathbb{R}^{3}\times\mathbb{R}^{3}}d\lambda d\mu\left(\mathcal{P}\left(\frac{\hbar \lambda\cdot\mu}{2}\right)\mathcal{K}\left(\frac{\pi k l}{s+1}\right)\right)^{-1}\exp\{i(\lambda\cdot p +\mu\cdot q)\}
$$
\begin{equation}\label{31}
\exp\left\{ i\frac{2\pi}{s+1}(km+ln)\right\}(\tilde{\tilde{f}}\boxtimes\tilde{\tilde{g}})(\lambda,\mu,k,l)
\end{equation}
where $(\tilde{\tilde{f}}\boxtimes\tilde{\tilde{g}})(\lambda,\mu,k,l)$ is given by (\ref{224}) with (\ref{225}).

Employing (\ref{238}) and (\ref{231b}) one can equivalently write (\ref{31})  as
$$
(f*g)(p,q,\phi_{m},n)=\frac{1}{\hbar^{6}(2\pi)^{12}(s+1)^{4}}\sum_{k,l,m',n', m'',n'',m''',n'''=0}^{s}
\int_{\mathbb{R}^{3}\times ... \times\mathbb{R}^{3}}d\lambda d\mu dp' dq' dp'' dq'' dp''' dq'''
$$
$$
\left|\mathcal{P}\left(\frac{\hbar \lambda\cdot\mu}{2}\right)\mathcal{K}\left(\frac{\pi k l}{s+1}\right)\right|^{-2} \exp\{i[\lambda\cdot(p-p') +\mu\cdot (q-q')]\}
$$
$$
\exp\left\{i\frac{2\pi}{s+1}[k(m-m')+l(n-n')]\right\}f(p'',q'',\phi_{m''},n'')
$$
\begin{equation}\label{32}
\textrm{Tr}\left\{\widehat{\Omega}[\mathcal{P},\mathcal{K}](p',q',\phi_{m'},n')\,\widehat{\Omega}[\mathcal{P},\mathcal{K}](p'',q'',\phi_{m''},n'')\,
\widehat{\Omega}[\mathcal{P},\mathcal{K}](p''',q''',\phi_{m'''},n''')\right\}g(p''',q''',\phi_{m'''},n''')
\end{equation}
The formula (\ref{32}) simplifies considerably when the conditions (\ref{237}) are fulfilled. In that case some integrals and sums can be calculated
explicitly and finally (\ref{32}) takes the form
$$
(f*g)(p,q,\phi_{m},n)=\frac{1}{(2\pi\hbar)^{6}(s+1)^{2}}\sum_{m',n', m'',n''=0}^{s}
\int_{\mathbb{R}^{3}\times\mathbb{R}^{3}\times\mathbb{R}^{3}\times\mathbb{R}^{3}} dp' dq' dp'' dq'' f(p',q',\phi_{m'},n')
$$
\begin{equation}\label{33}
\textrm{Tr}\left\{\widehat{\Omega}[\mathcal{P},\mathcal{K}](p,q,\phi_{m},n)\widehat{\Omega}[\mathcal{P},\mathcal{K}](p',q',\phi_{m'},n')
\widehat{\Omega}[\mathcal{P},\mathcal{K}](p'',q'',\phi_{m''},n'')\right\}g(p'',q'',\phi_{m''},n'')
\end{equation}
The mapping
\begin{equation}\label{34}
  *:(f,g) \mapsto f*g
\end{equation}
is called the \textit{star product} ($*$-product) and it is an obvious generalization of the celebrated Moyal $*$-product \cite{Groenewold,Moyal,Bayen,Plebanski1,Plebanski2,Plebanski3,ZFC} and the star product for  the discrete case \cite{Watson,Chaturvedi,Wootters} on the case of Hilbert space (\ref{21}).

\section{Wigner function and the Liouville - von Neumann - Wigner equation}

If $\widehat{\rho}$ is a density operator of the quantum system then the average value of an observable $\widehat{f}$ in the state $\widehat{\rho}$ reads
\begin{equation}\label{41}
  \langle \widehat{f} \rangle = \textrm{Tr}\{\widehat{f}\,\widehat{\rho}\}
\end{equation}
Inserting (\ref{231b}) into (\ref{41}) one gets
\begin{equation}\label{42}
 \langle \widehat{f} \rangle=\sum_{m,n=0}^{s}\int_{\mathbb{R}^{3}\times\mathbb{R}^{3}}dp dq f(p,q,\phi_{m},n) \frac{1}{(2\pi\hbar)^{3}(s+1)} \textrm{Tr}\left\{\widehat{\rho}\,\widehat{\Omega}[\mathcal{P},\mathcal{K}](p,q,\phi_{m},n)\right\}
\end{equation}
Hence, we define the \textit{Wigner function of the state} $\widehat{\rho}$ \textit{for the kernels} $(\mathcal{P},\mathcal{K})$ as
\begin{equation}\label{43}
  \rho_{W}[\mathcal{P},\mathcal{K}](p,q,\phi_{m},n):= \frac{1}{(2\pi\hbar)^{3}(s+1)} \textrm{Tr}\left\{\widehat{\rho}\, \widehat{\Omega}[\mathcal{P},\mathcal{K}](p,q,\phi_{m},n)\right\}
\end{equation}
Consequently, Eq. (\ref{42}) can be rewritten in the form
\begin{equation}\label{44}
  \langle \widehat{f} \rangle=\sum_{m,n=0}^{s}\int_{\mathbb{R}^{3}\times\mathbb{R}^{3}}dp dq f(p,q,\phi_{m},n) \rho_{W}[\mathcal{P},\mathcal{K}](p,q,\phi_{m},n)
\end{equation}
Employing the formulae (\ref{232}) and (\ref{235a}) one quickly finds that (\ref{43}) gives
$$
\rho_{W}[\mathcal{P},\mathcal{K}](p,q,\phi_{m},n):= \frac{1}{(2\pi)^{6}(s+1)^{2}}\sum_{k,l=0}^{s} \int_{\mathbb{R}^{3}\times\mathbb{R}^{3}}d\lambda d\mu \mathcal{P}^{*}\left(\frac{\hbar \lambda\cdot\mu}{2}\right)\mathcal{K}^{*}\left(\frac{\pi k l}{s+1}\right)
$$
\begin{equation}\label{45}
\exp\{i(\lambda\cdot p +\mu\cdot q)\}\exp\left\{ i\frac{2\pi}{s+1}(km+ln)\right\}
\textrm{Tr}\left\{\widehat{\rho}\,\widehat{\mathcal{U}}^{\dag}(\lambda,\mu)\widehat{\mathcal{D}}^{\dag}(k,l)\right\}
\end{equation}
The inverse relation to (\ref{43}) reads
$$
\widehat{\rho}=\frac{1}{(2\pi)^{6}(s+1)^{2}}\sum_{k,l,m,n,m',n'=0}^{s} \int_{\mathbb{R}^{3}\times ... \times\mathbb{R}^{3}}d\lambda d\mu dp dq dp' dq'
\left|\mathcal{P}\left(\frac{\hbar \lambda\cdot\mu}{2}\right)\mathcal{K}\left(\frac{\pi k l}{s+1}\right)\right|^{-2}
$$
$$
\exp\{i[\lambda\cdot(p-p')+\mu\cdot(q-q')]\}\exp\left\{i\frac{2\pi}{s+1}[k(m-m')+l(n-n')]\right\}
$$
\begin{equation}\label{46}
\rho_{W}[\mathcal{P},\mathcal{K}](p,q,\phi_{m},n)\widehat{\Omega}[\mathcal{P},\mathcal{K}](p',q',\phi_{m'},n')
\end{equation}
If the conditions (\ref{237}) are fulfilled then (\ref{46}) reduces to a simple relation
\begin{equation}\label{47}
  \widehat{\rho}=\sum_{m,n=0}^{s}\int_{\mathbb{R}^{3}\times\mathbb{R}^{3}}dp dq \rho_{W}[\mathcal{P},\mathcal{K}](p,q,\phi_{m},n)\widehat{\Omega}[\mathcal{P},\mathcal{K}](p,q,\phi_{m},n)
\end{equation}
One can easily check that the Wigner function defined by (\ref{43}) satisfies the usual requests imposed on Wigner functions:
\begin{enumerate}[(1)]
\item It is a real function
\begin{equation}\label{48}
  \rho_{W}^{*}[\mathcal{P},\mathcal{K}]=\rho_{W}[\mathcal{P},\mathcal{K}]
\end{equation}
\item Its trace is equal to one
\begin{equation}\label{49}
  \sum_{m,n=0}^{s}\int_{\mathbb{R}^{3}\times\mathbb{R}^{3}}dp dq \rho_{W}[\mathcal{P},\mathcal{K}](p,q,\phi_{m},n)=\textrm{Tr}\{\widehat{\rho}\}=1
\end{equation}
\item It gives the marginal distributions
\begin{subequations}\label{410}
\begin{align}
  \sum_{m,n=0}^{s}\int_{\mathbb{R}^{3}}dp \rho_{W}[\mathcal{P},\mathcal{K}](p,q,\phi_{m},n) ={}& \textrm{Tr}\{\widehat{\rho}|q\rangle\langle q|\} \label{410a}\\
  \sum_{m,n=0}^{s}\int_{\mathbb{R}^{3}}dq \rho_{W}[\mathcal{P},\mathcal{K}](p,q,\phi_{m},n) ={}& \textrm{Tr}\{\widehat{\rho}|p\rangle\langle p|\} \label{410b}\\
  \sum_{m=0}^{s}\int_{\mathbb{R}^{3}\times\mathbb{R}^{3}}dp dq \rho_{W}[\mathcal{P},\mathcal{K}](p,q,\phi_{m},n) ={}& \textrm{Tr}\{\widehat{\rho}|n\rangle\langle n|\} \label{410c} \\
  \sum_{n=0}^{s}\int_{\mathbb{R}^{3}\times\mathbb{R}^{3}}dp dq \rho_{W}[\mathcal{P},\mathcal{K}](p,q,\phi_{m},n) ={}& \textrm{Tr}\{\widehat{\rho}|\phi_{m}\rangle\langle \phi_{m}|\} \label{410d}
\end{align}
\end{subequations}
(where $|q\rangle\langle q|:=|q\rangle\langle q|\otimes \widehat{1}$, $|\phi_{m}\rangle\langle \phi_{m}|:= \widehat{1}\otimes |\phi_{m}\rangle\langle \phi_{m}|$, etc.).
\end{enumerate}
In the end we are considering the evolution equation for the Wigner function. Let the density operator depend on $t$, $\widehat{\rho}=\widehat{\rho}(t)$. Then the evolution equation for $\widehat{\rho}(t)$ is the \textit{Liouville-von Neumann equation}
\begin{equation}\label{411}
  \frac{\partial \widehat{\rho}}{\partial t} + \frac{1}{i \hbar}\left[\widehat{\rho}, \widehat{H}\right]=0
\end{equation}
where $\widehat{H}$ is the Hamilton operator.
\\
If the kernels $(\mathcal{P},\mathcal{K})$ do not satisfy the conditions (\ref{237}) then the respective evolution equation for the Wigner function $\rho_{W}[\mathcal{P},\mathcal{K}]$ is rather involved. So, in the general case we proceed as follows. We first define $\tilde{\tilde{\rho}}(\lambda,\mu,k,l)$ and $\tilde{\tilde{H}}(\lambda,\mu,k,l)$ according to the prescription (\ref{222}) i.e.
\begin{subequations}\label{412}
\begin{align}
  \tilde{\tilde{\rho}}(\lambda,\mu,k,l) ={}& \left(\frac{\hbar}{2\pi}\right)^{3}(s+1)^{-1} \textrm{Tr}\left\{\widehat{\rho}\,\widehat{\mathcal{U}}^{\dag}(\lambda,\mu)\widehat{\mathcal{D}}^{\dag}(k,l)\right\} \label{412a}\\
  \tilde{\tilde{H}}(\lambda,\mu,k,l) ={}& \left(\frac{\hbar}{2\pi}\right)^{3}(s+1)^{-1} \textrm{Tr}\left\{\widehat{H}\,\widehat{\mathcal{U}}^{\dag}(\lambda,\mu)\widehat{\mathcal{D}}^{\dag}(k,l)\right\} \label{412b}
\end{align}
\end{subequations}
Inserting (\ref{412a}) into (\ref{45}) one quickly gets
$$
\rho_{W}[\mathcal{P},\mathcal{K}](p,q,\phi_{m},n)=\frac{1}{(2\pi\hbar)^{3}(s+1)}\sum_{k,l,=0}^{s}\int_{\mathbb{R}^{3}\times\mathbb{R}^{3}}d\la d\mu
\mathcal{P}^{*}\left( \frac{\hbar\la \cdot \mu}{2}\right)\mathcal{K}^{*}\left( \frac{\pi k l}{s+1}\right)
$$
\begin{equation}\label{413}
\exp\{i(\la\cdot p+\mu\cdot q)\}\exp\left\{i\frac{2\pi}{s+1}(km+ln)\right\}\tilde{\tilde{\rho}}(\la,\mu,k,l).
\end{equation}
Therefore, the function $\tilde{\tilde{\rho}}$ determines uniquely $\rho_{W}[\mathcal{P},\mathcal{K}]$ according to (\ref{413}). Considering $\tilde{\tilde{\rho}}$ as dependent on time and using (\ref{411}), (\ref{224}) with (\ref{225}) we find the evolution equation for $\tilde{\tilde{\rho}}=\tilde{\tilde{\rho}}(\la,\mu,k,l;t)$
\begin{equation}\label{414}
\frac{\p\tilde{\tilde{\rho}}}{\p t}+\frac{1}{i\hbar}\left(\tilde{\tilde{\rho}}\boxtimes\tilde{\tilde{H}}-\tilde{\tilde{H}}\boxtimes\tilde{\tilde{\rho}}\right)=0.
\end{equation}
Finally, note that if the conditions (\ref{237}) are satisfied then the evolution equation for the Wigner function $\rho_{W}[\mathcal{P},\mathcal{K}](p,q,\phi_{m},n;t)$ reads
\begin{equation}
\label{415}
\frac{\p\rho_{W}[\mathcal{P},\mathcal{K}]}{\p t}+\frac{1}{i\hbar}\bigg( \rho_{W}[\mathcal{P},\mathcal{K}]* H - H * \rho_{W}[\mathcal{P},\mathcal{K}]\bigg)=0,
\end{equation}
where the Hamiltonian $H=H(p,q,\phi_{m},n)$ is defined according to the rule (\ref{239})
\begin{equation}
\label{416}
H(p,q,\phi_{m},n)= \textrm{Tr}\left\{\widehat{H}\widehat{\Omega}[\mathcal{P},\mathcal{K}](p,q,\phi_{m},n)\right\},
\end{equation}
and the $\ast$-product is given by (\ref{33}).
\\
Eq. (\ref{415}) (or Eq. (\ref{414})) will be called the \textit{Liouville-von Neumann-Wigner equation}.
\section{Remarks on the choice of kernels $(\mathcal{P},\mathcal{K})$}
As we have shown in Section 2 the kernels $\mathcal{P}$ and $\mathcal{K}$ should satisfy the conditions (\ref{233}) and (\ref{234}). We can then observe that many formulas take the much simpler forms when the conditions (\ref{237}) are also fulfilled. One quickly finds that assuming
\begin{equation}\label{51}
\mathcal{P}\left(\frac{\hbar\la \cdot \mu}{2}\right)=1, \hspace{1cm} \forall \la, \mu\in\mathbb{R}^{3},
\end{equation}
we satisfy the requirements (\ref{233}), (\ref{234a}) and (\ref{237}) for the kernel $\mathcal{P}$. As is well known \cite{Weyl1,Weyl2,Cohen,Tatarskij,Plebanski2} the choice (\ref{51}) realizes the \textit{Weyl ordering} of operators in $L^{2}(\mathbb{R}^{3})$.
\\
However, at the analogous assumption about the kernel $\mathcal{K}$, i.e. $\mathcal{K}\left(\frac{\pi  k l}{s+1}\right)=1$ $\forall$ $0 \leq k,l\leq s$, the relations (\ref{234b}) are not fulfilled for every $(k, l)$. Therefore one is forced to look for other options. The simplest one seems to be $\mathcal{K}(0)= 1$ and $\mathcal{K}\left(\frac{\pi k l}{s+1}\right)=\pm1$ for $kl\neq0$. This can be indeed realized as follows:
\begin{enumerate}[(a)]
\item If $s+1=$ odd number one puts
\begin{equation}\label{52}
\mathcal{K}\left(\frac{\pi k l}{s+1}\right)=(-1)^{k l}
\end{equation}
\item If $s+1$ is even  and $\frac{s+1}{2}$ is an odd number one assumes
\begin{equation}\label{53}
\mathcal{K}\left(\frac{\pi k l}{s+1}\right)=\left\{
\begin{array}{lll}
-1 & \textup{for} & k l=\textup{odd number}\\
-1 & \textup{for} & k l=2p, \,\,\,  p=\textup{odd number}\\
+1 & \textup{for} & k l=4r
\end{array}\right.
\end{equation}
\item If $s+1$  and $\frac{s+1}{2}$ are even numbers then we have not been able still to find the concise formula but one can realize the option: $\mathcal{K}\left(\frac{\pi k l}{s+1}\right)=\pm1$ for $k l\neq0$, $\mathcal{K}(0)=1$, by using the relation (\ref{234b}) step by step. We present this procedure for the case of $s+1=4$ in the next paper.
\end{enumerate}
The choice (a) was considered previously by several authors \cite{Wootters,Cohendet,Takami,Vaccaro,Leonhardt}.
\\
In the recent work \cite{Przan1} we have found the kernel $\mathcal{K}$ which satisfies all desired conditions (\ref{229}), (\ref{233}) and (\ref{234b}) but does not have the property (\ref{237}).
\\
This $\mathcal{K}$ reads
\begin{equation}\label{54}
\mathcal{K}\left(\frac{\pi k l}{s+1}\right)=\frac{\cos\left(\frac{\pi k l}{s+1}+\epsilon_{s}\right)}{\cos(\epsilon_{s})},
\end{equation}
where $\epsilon_{s}$ is taken so as to ensure that (\ref{229}) holds. In particular if $s+1=$ odd number, then one can put $\epsilon_{s}=0$. In that case we have
\begin{equation}\label{55}
\mathcal{K}\left(\frac{\pi k l}{s+1}\right)=\cos\left(\frac{\pi k l}{s+1}\right), \,\,\, s+1= \textrm{odd\; number},
\end{equation}
and such a kernel leads to the \textit{symmetric ordering} of operators in $\mathcal{H}^{(s+1)}$.
\\
When $s+1=$ even number, we must take $\epsilon_{s}\neq 0$. Nevertheless $|\epsilon_{s}|$ can be chosen arbitrary small (but $\neq0$). As has been shown in \cite{Przan1} with such a choice of $\epsilon_{s}$ the kernel (\ref{54}) realizes the \textit{almost symmetric ordering} of operators in $\mathcal{H}^{(s+1)}$. The kernel (\ref{54}) (all the more (\ref{55})) has a nice and useful feature. Namely the Stratonovich-Weyl quantizer defined by this kernel in $\mathcal{H}^{(s+1)}$ has relatively a simple form. Consequently, also the respective Wigner function takes a transparent form.
\\
Remembering all that we assume that the kernel $\mathcal{P}$ is given by (\ref{51}) and $\mathcal{K}$ is defined by (\ref{54}). Under these assumptions, from (\ref{232}) with (\ref{212}) and (\ref{215}) one gets the corresponding Stratonovich-Weyl quantizer as
$$
\widehat{\Omega}(p,q,\phi_{m},n)=\int_{\mathbb{R}^{3}}d\xi\exp\left\{i\frac{\xi \cdot p}{\hbar}\right\}\left|q+\frac{\xi}{2}\right\rangle\left\langle q-\frac{\xi}{2}\right|
$$
$$
\otimes\frac{s+1}{2\cos(\epsilon_{s})}\bigg[ \exp\{-i\epsilon_{s}\}|n\rangle\langle n|\phi_{m}\rangle\langle\phi_{m}|+ \textrm{Hermitian conjugation}\bigg]
$$
$$
=\frac{1}{2\cos(\epsilon_{s})}\bigg\{\exp\{-in\phi_{m}\}\sum_{n'=0}^{s}\int_{\mathbb{R}^{3}}d\xi \exp\left\{i\frac{\xi \cdot p}{\hbar}\right\}
\exp\{i(n'\phi_{m}+\epsilon_{s})\}
$$
\begin{equation} \label{56}
\left|q+\frac{\xi}{2},n'\right\rangle \left\langle q-\frac{\xi}{2},n\right| + \textrm{Hermitian conjugation}\bigg\},
\end{equation}
where $d\xi:=d\xi_1 d\xi_2 d\xi_3$
(see Eq. (2.13) in \cite{Plebanski3} and Eq. (5.2) in \cite{Przan1}).
\\
(\textit{Remark}: In (\ref{56}) and also further on we omit the symbol $[\mathcal{P},\mathcal{K}]$ for the kernels given by (\ref{51}) and (\ref{54})).
\\
Inserting (\ref{56}) into (\ref{43}) one quickly gets the Wigner function
\begin{eqnarray}\label{57}
\rho_{W}(p,q,\phi_{m},n)=\frac{1}{(2\pi\hbar)^{3}(s+1)\cos(\epsilon_{s})}\Re\bigg[\exp\{i(\epsilon_{s}-n\phi_{m})\}\nonumber\\
\left.\sum_{n'=0}^{s}\int_{\mathbb{R}^{3}}d\xi\exp\left\{i\frac{\xi\cdot p}{\hbar}\right\}\exp\{in'\phi_{m}\}\left\langle q-\frac{\xi}{2},n\right|\widehat{\rho}\left|q+\frac{\xi}{2},n'\right\rangle\right.\bigg].
\end{eqnarray}
In particular, for the pure state
\begin{equation}\label{58}
\widehat{\rho}=|\psi\rangle\langle\psi|, \hspace{1cm}\langle\psi|\psi\rangle=1,
\end{equation}
Eq. (\ref{57}) gives
\begin{eqnarray}\label{59}
\rho_{W}(p,q,\phi_{m},n)=\frac{1}{(2\pi\hbar)^{3}(s+1)\cos(\epsilon_{s})}\Re\bigg[\exp\{i(\epsilon_{s}-n\phi_{m})\}\nonumber\\
\sum_{n'=0}^{s}\int_{\mathbb{R}^{3}}d\xi\exp\left\{i\frac{\xi\cdot p}{\hbar}\right\}\exp\{in'\phi_{m}\}\psi_{n'}^{*}\left(q+\frac{\xi}{2}\right)\psi_{n}\left(q-\frac{\xi}{2}\right)\bigg],
\end{eqnarray}
where
\begin{equation}\label{510}
\psi_{j}(q):=\langle q,j|\psi\rangle, \hspace{1cm} j=0,...,s; \,\,\, q\in\mathbb{R}^{3}.
\end{equation}
Formulas (\ref{57}) and (\ref{59}) are the main results of our paper. They define the Wigner function for nonrelativistic or relativistic quantum particle described by the density operator $\widehat{\rho}$ in the Hilbert space $L^{2}(\mathbb{R}^3)\times\mathcal{H}^{(s+1)}$. This can be immediately extended to the case $L^{2}(\mathbb{R}^{3} \times \mathbb{R}^{3}\times  \cdots  \times \mathbb{R}^{3})\times\mathcal{H}^{(s+1)}$.
\\
Note finally that having given the Wigner function $\rho_{W}(p,q,\phi_{m},n)$ one can find the respective Wigner function $\rho_{W}[\mathcal{P}',\mathcal{K}'](p,q,\phi_{m},n)$ for any kernels $(\mathcal{P}',\mathcal{K}')$ as
\begin{eqnarray}\label{511}
\rho_{W}[\mathcal{P}',\mathcal{K}'](p,q,\phi_{m},n)=\frac{1}{(2\pi)^{6}(s+1)^{2}}\sum_{k,l,m',n'=0}^{s}\int_{\mathbb{R}^{3}\times\mathbb{R}^{3}\times\mathbb{R}^{3}\times\mathbb{R}^{3}}d\la d\mu dp'dq' \nonumber \\
\mathcal{P}'\left(\frac{\hbar\la\cdot\mu}{2}\right)\mathcal{K}'\left(\frac{\pi k l}{s+1}\right)\left(\frac{\cos\left(\frac{\pi k l}
{s+1}+\epsilon_{s}\right)}{\cos(\epsilon_{s})}\right)^{-1}\exp\{i[\la\cdot(p'-p)+\mu\cdot(q'-q)]\} \nonumber \\
\exp\left\{i\frac{2\pi}{s+1}[k(m'-m)+l(n'-n)]\right\}\rho_{W}(p',q',\phi_{m'},n').
\end{eqnarray}
Then according to (\ref{413}) the Wigner function $\rho_{W}$ is related to $\tilde{\tilde{\rho}}$, defined by (\ref{412a}), as follows
\begin{eqnarray}\label{512}
\rho_{W}(p,q,\phi_{m},n)=\frac{1}{(2\pi\hbar)^{3}(s+1)\cos(\epsilon_{s})}\sum_{k,l=0}^{s}\int_{\mathbb{R}^{3}\times\mathbb{R}^{3}}d\lambda d\mu \cos\left(\frac{\pi k l}{s+1}+\epsilon_{s}\right)\nonumber\\
\exp\{i(\la\cdot p+\mu\cdot q)\}\exp\left\{i\frac{2\pi}{s+1}(km+ln)\right\}\tilde{\tilde{\rho}}(\lambda,\mu,k,l).
\end{eqnarray}
For the pure state (\ref{58}) $\tilde{\tilde{\rho}}$ reads (use (\ref{212}) and (\ref{215}))
\begin{eqnarray}\label{513}
\tilde{\tilde{\rho}}(\lambda,\mu,k,l)=\left(\frac{\hbar}{2\pi}\right)^{3}(s+1)^{-1}\exp\left\{-i\frac{\pi k l}{s+1}\right\}\sum_{n=0}^{s}\int_{\mathbb{R}^{3}}dq\exp\{-i\mu\cdot q\}\nonumber\\
\exp\left\{-i\frac{2\pi ln}{s+1}\right\}\psi_{n+k mod(s+1)}^{*}\left(q+\frac{\hbar\la}{2}\right)\psi_{n}\left(q-\frac{\hbar\la}{2}\right).
\end{eqnarray}

\section{Examples}
This section contains detailed analysis of the proposed phase space formalism
to the well known problems:   motion of   spin $\frac{1}{2}$ nonrelativistic particle in a homogeneous magnetic field and
magnetic resonance for a spin $\frac{1}{2}$ uncharged nonrelativistic particle.

\subsection{Spin $\frac{1}{2}$ nonrelativistic particle in a homogeneous magnetic field. Landau levels.}
Consider a spin $\frac{1}{2}$ nonrelativistic particle in a classical electromagnetic field described by the vector potential $A=(A_{1},A_{2},A_{3})$ and the scalar potential $A_{0}$. The Hamilton operator, called the \textit{Pauli Hamiltonian}, reads now
\begin{equation}\label{61}
\widehat{H}_{P}=\frac{1}{2m_{0}}\sum_{j=1}^{3}\left(\widehat{p}_j-\frac{e_{0}}{c}\widehat{A}_j\right)^{2}+e_{0}\widehat{A}_{0}-\frac{e_{0}\hbar}{2m_{0}c}\widehat{\sigma}\cdot \widehat{B},
\end{equation}
where $m_{0}$ and $e_{0}$ stand for the mass and the charge of particle, respectively, $c$ denotes the speed of light, $B=(B_1, B_2, B_3)$ is the magnetic induction and $\widehat{\sigma}=(\widehat{\sigma}_{1},\widehat{\sigma}_{2},\widehat{\sigma}_{3})$ defines the spin operator by $\widehat{S}=\frac{\hbar}{2}\widehat{\sigma}$. Then the \textit{Pauli evolution equation} for a pure state $|\psi(t)\rangle$ is
\begin{equation}\label{62}
i\hbar\frac{\p|\psi(t)\rangle}{\p t}=\widehat{H}_{P}|\psi(t)\rangle,
\end{equation}
and the corresponding \textit{Pauli-Liouville-von Neumann equation} for the density operator reads (see \ref{411})
\begin{equation}\label{63}
\frac{\p \widehat{\rho}}{\p t}+\frac{1}{i\hbar}\left[\widehat{\rho},\widehat{H}_{P}\right]=0.
\end{equation}
In the present case the Hilbert space $\mathcal{H}^{(s+1)}$ given in (\ref{21}) is 2-dimensional, $s+1=2$, and the Pauli spin operator $\widehat{S}$ acts in that space. We choose the orthonormal basis $\{|n\rangle\}_{n=0,1}$ in $\mathcal{H}^{(2)}$ in a standard way i.e. so that
\begin{equation}\label{64}
\widehat{\sigma}_{3}|0\rangle=1\cdot|0\rangle,\hspace{1cm}\widehat{\sigma}_{3}|1\rangle=-1\cdot|1\rangle.
\end{equation}
In this basis the operator $\widehat{\sigma}$ is represented by the \textit{Pauli matrices}
\begin{equation}\label{65}
\sigma=(\sigma_{1},\sigma_{2},\sigma_{3}),\hspace{1cm}\sigma_{1}=\left(
\begin{array}{cc}
0&1\\
1&0
\end{array}
\right),\hspace{1cm}\sigma_{2}=\left(
\begin{array}{cc}
0&-i\\
i&0
\end{array}
\right),\hspace{1cm}\sigma_{1}=\left(
\begin{array}{cc}
1&0\\
0&-1
\end{array}
\right).
\end{equation}
Proceeding to the phase space representation of our quantum system we first note that now the phase space is
\begin{equation}\label{66}
\Gamma=\{(p,q,\phi_{m},n)\}=\mathbb{R}^{3}\times\mathbb{R}^{3}\times\{(\phi_{m},n)\}_{m,n=0,1}.
\end{equation}
Then, inserting $s+1=2$ into Eq. (\ref{54}) and assuming $\epsilon_{s}=\frac{\pi}{4}$ one quickly gets a simple kernel
\begin{equation}\label{67}
\mathcal{K}\left(\frac{\pi k l}{2}\right)=(-1)^{k l}, \qquad k,l=0,1
\end{equation}
(Note that the same kernel (\ref{67}) can be  obtained from (\ref{53}) with $s=1$). Assuming also that (\ref{51}) holds true we easily find the
Stratonovich-Weyl quantizer from (\ref{56}) as
\begin{multline}\label{68}
\widehat{\Omega}(p,q,\phi_{m},n)=\int_{\mathbb{R}^{3}}d\xi\exp\left\{i\frac{\xi\cdot p}{\hbar}\right\}\left|q+\frac{\xi}{2}\right\rangle\left\langle q-\frac{\xi}{2}\right| \\
\otimes\bigg[(1-i)|n\rangle\langle n|\phi_{m}\rangle\langle\phi_{m}|+(1+i)|\phi_{m}\rangle\langle\phi_{m}|n\rangle\langle n|\bigg]  \\
=\frac{1-i}{\sqrt{2}}\exp\{in\phi_{m}\}\int_{\mathbb{R}^{3}}d\xi\exp\left\{i\frac{\xi\cdot p}{\hbar}\right\}
\left|q+\frac{\xi}{2},n\right\rangle\left\langle q-\frac{\xi}{2},\phi_{m}\right|+\textup{Hermitian conjugation}.
\end{multline}
Straightforward calculations show that (\ref{68}) can be also rewritten in the following useful form
\begin{equation}\label{69}
\widehat{\Omega}(p,q,\phi_{m},n)=\int_{\mathbb{R}^{3}}d\xi\exp\left\{i\frac{\xi\cdot p}{\hbar}\right\}\left|q+\frac{\xi}{2}\right\rangle\left\langle q-\frac{\xi}{2}\right|
\otimes\frac{1}{2}\bigg[1+(-1)^{m}\widehat{\sigma}_{1}+(-1)^{m+n}\widehat{\sigma}_{2}+(-1)^{n}\widehat{\sigma}_{3}\bigg]
\end{equation}
(The ``discrete part" of the $\widehat{\Omega}$ i.e. $\frac{1}{2}[1+(-1)^{m}\widehat{\sigma}_{1}+(-1)^{m+n}\widehat{\sigma}_{2}+(-1)^{n}\widehat{\sigma}_{3}]$ was
found by W. K. Wootters \cite{Wootters} and R. Feynman \cite{Feynman}, then also by S. Chaturvedi et al \cite{Chaturvedi} and very recently by us  \cite{Przan1}).
\\
From (\ref{33}) we conclude that in order to get the $\ast$-product one needs the explicit expression for
$\textrm{Tr} \{\widehat{\Omega}(p,q,\phi_{m},n)\widehat{\Omega}(p',q',\phi_{m'},n')\widehat{\Omega}(p'',q'',\phi_{m''},n'')\}$.
\\
Employing (\ref{69}) and performing some simple manipulations we obtain
\begin{multline}\label{610}
\textrm{Tr}\{\widehat{\Omega}(p,q,\phi_{m},n)\widehat{\Omega}(p',q',\phi_{m'},n')\widehat{\Omega}(p'',q'',\phi_{m''},n'')\}= \\
64\exp\left\{\frac{2i}{\hbar}[(q-q')\cdot(p-p'')-(q-q'')\cdot(p-p')]\right\}  \\ \cdot
\frac{1}{4}\left\{(1+(-1)^{m'+m''})(1+(-1)^{n'+n''})+(-1)^{m}((-1)^{m'}+(-1)^{m''})+(-1)^{m+n}((-1)^{m'+n'}+(-1)^{m''+n''})\right.\\
+(-1)^{n}((-1)^{n'}+(-1)^{n''})+i\left[(-1)^{m}(-1)^{n'+n''}((-1)^{m'}-(-1)^{m''})\right.\\
\left.\left.+(-1)^{m+n}((-1)^{m''+n'}-(-1)^{m'+n''})+(-1)^{n}(-1)^{m'+m''}((-1)^{n''}-(-1)^{n'})\right]\right\}.
\end{multline}
The ``continuous part" of the formula (\ref{610}) i.e. $ 64\exp\left\{\frac{2i}{\hbar}[(q-q')\cdot(p-p'')-(q-q'')\cdot(p-p')]\right\}$ leads to the
celebrated \textit{Moyal $*$-product} for the functions on $\mathbb{R}^{3}\times\mathbb{R}^{3}$. Performing some standard manipulations one quickly finds that
if the functions $f$ and $g$ are independent of $(\phi_{m},n)$ i.e. $f=f(p,q)$ and $g=g(p,q)$ then $f*g$ can be written as a formal series in $\hbar$
\begin{equation}\label{611}
(f \ast g)(p,q)=f(p,q)\exp\left\{\frac{i\hbar}{2}\overleftrightarrow{\mathcal{P}}\right\}g(p,q),
\end{equation}
where $\overleftrightarrow{\mathcal{P}}$ denotes the Poisson operator
\begin{equation}\label{612}
\overleftrightarrow{\mathcal{P}}:=\sum_{j=1}^{3}\left(\frac{\overleftarrow{\p}}{\p q_{j}}\frac{\overrightarrow{\p}}{\p p_{j}}-
\frac{\overleftarrow{\p}}{\p p_{j}}\frac{\overrightarrow{\p}}{\p q_{j}}\right).
\end{equation}
Hence, in the general case we get
\begin{multline}\label{613}
(f * g)(p,q,\phi_{m},n)=\frac{1}{16}\sum_{m',n',m'',n''=0}^{1}f(p,q,\phi_{m'},n')\exp\left\{\frac{i\hbar}{2}\overleftrightarrow{\mathcal{P}}\right\}g(p,q,\phi_{m}'',n'')\\
\left\{(1+(-1)^{m'+m''})(1+(-1)^{n'+n''})+(-1)^{m}((-1)^{m'}+(-1)^{m''})+(-1)^{m+n}((-1)^{m'+n'}+(-1)^{m''+n''})\right.\\
+(-1)^{n}((-1)^{n'}+(-1)^{n''})+i\left[(-1)^{m}(-1)^{n'+n''}((-1)^{m'}-(-1)^{m''})\right.\\
\left.\left.+(-1)^{m+n}((-1)^{m''+n'}-(-1)^{m'+n''})+(-1)^{n}(-1)^{m'+m''}((-1)^{n''}-(-1)^{n'})\right]\right\}.
\end{multline}
Then the \textit{$*$-bracket}
\begin{equation}\label{614}
[f,g]_{*}:=f * g-g * f,
\end{equation}
reads
\begin{multline}\label{615}
[f,g]_{*}(p,q,\phi_{m},n)=\frac{i}{8}\sum_{m',n',m'',n''=0}^{1}f(p,q,\phi_{m'},n')\left\{\left[(1+(-1)^{m'+m''})(1+(-1)^{n'+n''})\right.\right.\\
+(-1)^{m}((-1)^{m'}+(-1)^{m''})+(-1)^{m+n}((-1)^{m'+n'}+(-1)^{m''+n''})\\
\left.+(-1)^{n}((-1)^{n'}+(-1)^{n''})\right]\sin\left(\frac{\hbar}{2}\overleftrightarrow{\mathcal{P}}\right)
+\left[(-1)^{m}(-1)^{n'+n''}((-1)^{m'}-(-1)^{m''})\right.\\
\left.+(-1)^{m+n}((-1)^{m''+n'}-(-1)^{m'+n''})+(-1)^{n}(-1)^{m'+m''}((-1)^{n''}-(-1)^{n'})\right]\\
\left.\cos\left(\frac{\hbar}{2}\overleftrightarrow{\mathcal{P}}\right)\right\}g(p,q,\phi_{m''},n'').
\end{multline}
From (\ref{416}) with the Hamilton operator given by the Pauli Hamiltonian (\ref{61}) and the Stratonovich-Weyl quantizer defined by (\ref{69}) one easily gets
\begin{multline}\label{616}
H_{P}(p,q,\phi_{m},n)=\frac{1}{2m_{0}} \sum_{j=1}^3 \left(p_j-\frac{e_{0}}{c}A_j\right)\ast\left(p_j-\frac{e_{0}}{c}A_j\right)+e_{0}A_{0}\\
-\frac{e_{0}\hbar}{2m_{0}c}\bigg[(-1)^{m}B_{1}+(-1)^{m+n}B_{2}+(-1)^{n}B_{3}\bigg]\\
=\frac{1}{2m_{0}} \sum_{j=1}^3 \left(p_j-\frac{e_{0}}{c}A_j\right)^{2}+e_{0}A_{0}-\frac{e_{0}\hbar}{2m_{0}c}\bigg[(-1)^{m}B_{1}+(-1)^{m+n}B_{2}+(-1)^{n}B_{3}\bigg].
\end{multline}
Consider a simple case of the particle in homogeneous and time independent  magnetic field. We take the coordinates and the gauge so that (the \textit{Landau gauge} \cite{Landau1,Landau2,Tong,Murayama})
\begin{equation}\label{617}
  \begin{array}{ccc}
    B=(0,0,B_{3}), & \hspace{1cm} & B_{3}=\textrm{const.}>0, \\
    A_{0}=0, & \hspace{1cm} & A=(-q_{2}B_{3},0,0).
  \end{array}
\end{equation}
Inserting (\ref{617}) into (\ref{616}), and using then (\ref{615}) in (\ref{415}) with $H\rightarrow H_{P}$, after rather simple but tedious calculations one gets the
phase space image of the Pauli-Liouville-von Neumann equation (\ref{63}) in the form
\begin{multline}\label{618}
\frac{\p(\rho_{W})}{\p t}+\frac{1}{m_{0}}\left[\left(p_{1}+\frac{e_{0}B_{3}}{c}q_{2}\right)\left(\frac{\p}{\p q_{1}}-
\frac{e_{0}B_{3}}{c}\frac{\p}{\p p_{2}}\right)+p_{2}\frac{\p}{\p q_{2}}+p_{3}\frac{\p}{\p q_{3}}\right](\rho_{W})\\
-\frac{e_{0}B_{3}}{2m_{0}c}\left(
\begin{array}{cccc}
0&-1&0&1\\
1&0&-1&0\\
0&1&0&-1\\
-1&0&1&0
\end{array}
\right)\cdot(\rho_{W})=0,
\end{multline}
where $(\rho_{W})$ is the following one-column matrix
\begin{equation}\label{619}
(\rho_{W}):=\left(\begin{array}{c}
\rho_{W}(p,q,\phi_{0},0;t)\\
\rho_{W}(p,q,\phi_{0},1;t)\\
\rho_{W}(p,q,\phi_{1},0;t)\\
\rho_{W}(p,q,\phi_{1},1;t)
\end{array}\right).
\end{equation}
In particular, consider the case when the Wigner function $\rho_{W}=\rho_{W}(p,q,\phi_{m},n;t)$ corresponds to an eigenstate of $\widehat{\sigma}_{3}$. The eigenvalue equation reads
\begin{equation}\label{620}
\sigma_{3}(\phi_{m},n)*\rho_{W}(p,q,\phi_{m},n;t)=\la_0\rho_{W}(p,q,\phi_{m},n;t),
\end{equation}
where according to (\ref{239}) with (\ref{69})
\begin{equation}\label{621}
\sigma_{3}(\phi_{m},n) = \textrm{Tr}\{\widehat{\sigma}_{3}\widehat{\Omega}(p,q,\phi_{m},n)\}=(-1)^{n}.
\end{equation}
Using (\ref{613}) one can rewrite (\ref{620}) as
\begin{eqnarray}\label{622}
\rho_{W}(p,q,\phi_{0},0;t)+\rho_{W}(p,q,\phi_{1},0;t)+i\Big(\rho_{W}(p,q,\phi_{0},1;t)-\rho_{W}(p,q,\phi_{1},1;t)\Big)=2\la_0\rho_{W}(p,q,\phi_{0},0;t),\nonumber\\
\rho_{W}(p,q,\phi_{0},1;t)+\rho_{W}(p,q,\phi_{1},1;t)+i\Big(\rho_{W}(p,q,\phi_{0},0;t)-\rho_{W}(p,q,\phi_{1},0;t)\Big)=-2\la_0\rho_{W}(p,q,\phi_{0},1;t),\nonumber\\
\rho_{W}(p,q,\phi_{0},0;t)+\rho_{W}(p,q,\phi_{1},0;t)+i\Big(\rho_{W}(p,q,\phi_{1},1;t)-\rho_{W}(p,q,\phi_{0},1;t)\Big)=2\la_0\rho_{W}(p,q,\phi_{1},0;t),\nonumber\\
\rho_{W}(p,q,\phi_{0},1;t)+\rho_{W}(p,q,\phi_{1},1;t)+i\Big(\rho_{W}(p,q,\phi_{1},0;t)-\rho_{W}(p,q,\phi_{0},0;t)\Big)=-2\la_0\rho_{W}(p,q,\phi_{1},1;t),\nonumber\\
\end{eqnarray}
for real $\rho_{W}(p,q,\phi_{m},n;t)$.
\\
We quickly find the solutions of (\ref{622}) in the form:
\begin{subequations}\label{623}
\begin{align}
\la_0=1,\hspace{1cm}(\rho_{W})=\rho(p,q;t)\frac{1}{2}\left(
\begin{array}{c}
1\\
0\\
1\\
0
\end{array}
\right), \label{623a} \\
\intertext{or}
\la_0=-1,\hspace{1cm}(\rho_{W})=\rho(p,q;t)\frac{1}{2}\left(
\begin{array}{c}
0\\
1\\
0\\
1
\end{array}
\right), \label{623b}
\end{align}
\end{subequations}
where $\rho(p,q;t)$ is a real function.
\\
Inserting (\ref{623a}) or (\ref{623b}) into (\ref{618}) one gets the linear first-order PDE on $\rho(p,q;t)$
\begin{equation}\label{624}
\frac{\p\rho}{\p t}+\frac{1}{m_{0}}\left[\left(p_{1}+\frac{e_{0}B_{3}}{c}q_{2}\right)\left(\frac{\p}{\p q_{1}}-
\frac{e_{0}B_{3}}{c}\frac{\p}{\p p_{2}}\right)+p_{2}\frac{\p}{\p q_{2}}+p_{3}\frac{\p}{\p q_{3}}\right]\rho=0.
\end{equation}
Employing the standard method of the theory of linear first-order PDEs we find the general solution of Eq. (\ref{624}) as
\begin{eqnarray}\label{625}
\rho=\rho\bigg(p_{1},p_{2}+\frac{e_{0}B_{3}}{c}q_{1},p_{3},\frac{1}{2m_{0}}\left[\left(p_{1}+\frac{e_{0}B_{3}}{c}q_{2}\right)^{2}+p_{2}^{2}\right],\nonumber\\
\mp\frac{e_{0}B_{3}}{c}q_{3}+p_{3}\arcsin\left(\frac{p_{1}+\frac{e_{0}B_{3}}{c}q_{2}}{\sqrt{\left(p_{1}+\frac{e_{0}B_{3}}{c}q_{2}\right)^{2}+
p_{2}^{2}}}\right),q_{3}-\frac{p_{3}}{m_{0}}t\bigg),
\end{eqnarray}
with the upper sign ``-" corresponding to $p_{2}<0$ and the lower sign ``+" corresponding to $p_{2}>0$.
\\
This change of the sign at $p_{2}=0$ leads to a ``jump" of $\rho$ at $p_{2}=0$. Consequently, one concludes that the global general solution should be of the form
\begin{eqnarray}\label{626}
\rho=\rho\bigg(p_{1},p_{2}+\frac{e_{0}B_{3}}{c}q_{1},p_{3},\frac{1}{2m_{0}}\left[\left(p_{1}+\frac{e_{0}B_{3}}{c}q_{2}\right)^{2}+p_{2}^{2}\right],q_{3}-\frac{p_{3}}{m_{0}}t\bigg).
\end{eqnarray}
Assume that $\widehat{\rho}$ is not only an eigenstate of $\widehat{\sigma}_3$ but also  of the Pauli Hamiltonian $\widehat{H}_{P}$. Then $\frac{\p\widehat{\rho}}{\p t}=0\Rightarrow\frac{\p\rho_{W}}{\p t}=0$.
\\
The eigenvalue equation for $H_{P}$ in the phase space $\Gamma$ reads
\begin{equation}\label{627}
H_{P}*\rho_{W}=E\rho_{W}.
\end{equation}
Substituting $H_{P}$ given by (\ref{616}) with (\ref{617}) and $\rho_{W}$ defined by (\ref{623a}) or (\ref{623b}) into (\ref{627}) one has
\begin{equation}\label{628}
\frac{1}{2m_{0}}\left[\left(p_{1}+\frac{e_{0}B_{3}}{c}q_{2}\right)^{2}+p_{2}^{2}+p_{3}^{2}\right]*\rho-\frac{e_{0}\hbar B_{3}}{2m_{0}c}\la_0\rho=E\rho,\hspace{1cm}\la_0=\pm1.
\end{equation}
Inserting (\ref{626}) into (\ref{628}), remembering also that $\frac{\p\rho}{\p t}=0\Rightarrow\frac{\p\rho}{\p q_{3}}=0$ we arrive at the equation
\begin{multline}\label{629}
\frac{1}{2m_{0}}\left[\left(p_{1}+\frac{e_{0}B_{3}}{c}q_{2}\right)^{2}+p_{2}^{2}+p_{3}^{2}\right]\rho - \frac{\hbar^{2}}{8m_{0}}\left[\frac{\p^{2}\rho}{\p q_{1}^{2}}+\frac{\p^{2}\rho}{\p q_{2}^{2}}+\left(\frac{e_{0}B_{3}}{c}\right)^{2}\frac{\p^{2}\rho}{\p p_{2}^{2}}-2\frac{e_{0}B_{3}}{c}\frac{\p^{2}\rho}{\p p_{2}\p q_{1}}\right]\\
=\left(E+\frac{e_{0}\hbar B_{3}}{2m_{0}c}\la_0\right)\rho, \hspace{1cm}\la_0=\pm1,
\end{multline}
and finally, using again (\ref{626}), we obtain
\begin{equation}\label{630}
H'\rho-\frac{\hbar^{2}}{4}\omega_{0}^{2}\left(H'\frac{\p^{2}\rho}{\p H'^{2}}+\frac{\p\rho}{\p H'}\right)=
\left(E+\frac{e_{0}\hbar B_{3}}{2m_{0}c}\la_0-\frac{p_{3}^{2}}{2m_{0}}\right)\rho,
\end{equation}
where
\begin{equation}\label{631}
\omega_{0}:=\frac{|e_{0}|B_{3}}{m_{0}c},
\end{equation}
and
\begin{equation}\label{632}
H':=\frac{1}{2m_{0}}\left[\left(p_{1}+\frac{e_{0}B_{3}}{c}q_{2}\right)^{2}+p_{2}^{2}\right]=
\frac{p_{2}^{2}}{2m_{0}}+ \frac{1}{2}m_{0}\omega_{0}^{2}\left[q_{2}-\left(-\frac{cp_{1}}{e_{0}B_{3}}\right)\right]^{2}.
\end{equation}
Since
\begin{equation}\label{633}
[p_{1},H_{P}]_{\ast}=0=[p_{3},H_{P}]_{\ast},
\end{equation}
one can assume that $\rho$ is of the form
\begin{equation}\label{634}
\rho=\delta(p_{1}-p_{10})\delta(p_{3}-p_{30})\rho_{0}(H'_{0}),
\end{equation}
\[
p_{10}={\rm const.} \;\;\; {\rm and} \;\;\; p_{30}={\rm const.}
\]
where
\begin{equation}\label{635}
H_{0}':=\frac{p_{2}^{2}}{2m_{0}}+\frac{1}{2}m_{0}\omega_{0}^{2}\left[q_{2}-\left(-\frac{cp_{10}}{e_{0}B_{3}}\right)\right]^{2},
\end{equation}
and $\rho_{0}(H_{0}')$ satisfies the following linear second-order ODE
\begin{eqnarray}\label{636}
H_{0}'\rho_{0}-\frac{\hbar^{2}}{4}\left[\omega_{0}^{2}\left(H_{0}'\frac{d^{2}\rho_{0}}{dH_{0}'^{2}}+\frac{d\rho_{0}}{dH_{0}'}\right)\right]=\left(E+\frac{e_{0}\hbar B_{3}}{2m_{0}c}\la_0-\frac{p_{30}^{2}}{2m_{0}}\right)\rho_{0}, \hspace{1cm} \lambda_0=\pm1.
\end{eqnarray}
Comparing Eq. (\ref{636}) with Eq. (4.40) of  Ref. \cite{Tosiek} determining the Wigner function for the linear harmonic oscillator being in an energy eigenstate we quickly arrive at the conclusion that these two equations have the same form.
\\
Therefore, one immediately finds  possible values of $E$
\begin{equation}\label{637}
E_{N}=\hbar\omega_{0}\left(N+\frac{1}{2}-\frac{\lambda_0}{2}\cdot\textup{sgn}(e_{0})\right)+\frac{p_{30}^{2}}{2m_{0}},\hspace{1cm} \lambda_0=\pm1; \hspace{0.5cm} N=0,1,2,...
\end{equation}
Thus, of course, we recover the famous \textit{Landau levels} \cite{Landau1,Landau2,Tong,Murayama}. Then using again the results of \cite{Tosiek} (see also \cite{Bayen,Hirshfeld,Dito}) one gets $\rho_{0}$ for any $N=0,1,2,...$, $\rho_{0,N}$ as
\begin{equation}\label{638}
\rho_{0,N}=\frac{(-1)^{N}}{\pi\hbar}\exp\left\{-\frac{2H_{0}'}{\hbar\omega_{0}}\right\}L_{N}\left(\frac{4H_{0}'}{\hbar\omega_{0}}\right),
\end{equation}
where $L_{N}(x)$ denotes the Laguerre polynomial of the $N$th order $L_{N}(x)=\sum_{k=0}^{N}(-1)^{k}\binom{N}{N-k}\frac{x^{k}}{k!}$.
\\
Finally the Wigner functions satisfying (\ref{620}) and (\ref{627}) are given by
\begin{subequations}\label{639}
\begin{align}
(\rho_{W, N})=\frac{(-1)^{N}}{2\pi\hbar}\delta(p_{1}-p_{10})\delta(p_{3}-p_{30})\exp\left\{-\frac{2H_{0}'}{\hbar\omega_{0}}\right\}
L_{N}\left(\frac{4H_{0}'}{\hbar\omega_{0}}\right)\left(
\begin{array}{c}
1\\
0\\
1\\
0
\end{array}
\right),\label{639a} \\
\intertext{or}
(\rho_{W, N})=\frac{(-1)^{N}}{2\pi\hbar}\delta(p_{1}-p_{10})\delta(p_{3}-p_{30})\exp\left\{-\frac{2H_{0}'}{\hbar\omega_{0}}\right\}
L_{N}\left(\frac{4H_{0}'}{\hbar\omega_{0}}\right)\left(
\begin{array}{c}
0\\
1\\
0\\
1
\end{array}
\right),\label{639b}
\end{align}
\end{subequations}
with $N=0,1,2,...$.
\\
It is evident that the Wigner functions (\ref{639}) satisfy the eigenvalue equations for both $p_{1}$ and $p_{3}$
\begin{equation}\label{640}
p_{1}*\rho_{W, N} = p_{10}\rho_{W, N}, \hspace{1cm} p_{3}*\rho_{W, N}=p_{30}\rho_{W, N}, \hspace{1cm} N=0,1,2,...
\end{equation}
Note that analogous considerations but for a spinless particle have been done in \cite{Demircioglu} and for noncommutative space in \cite{Dayi}.

\subsection{Magnetic resonance}
In this subsection we study the magnetic resonance for a spin $\frac{1}{2}$ uncharged nonrelativistic particle equipped with a magnetic moment in terms of the Weyl-Wigner-Moyal formalism.
\\
The magnetic field has the form \cite{Rabi,Bialynicki,Levitt,LeBellac}
\begin{equation}\label{641}
B=(b\cos(\omega t),-b\sin(\omega t),B_{3}), \hspace{1cm} b,B_{3}=\textrm{const.}\,\, \textrm{ and } \omega = \textrm{const.}
\end{equation}
Inserting (\ref{641}) and assuming that
\begin{equation}\label{642}
e_{0}\rightarrow0,\hspace{1cm}\frac{e_{0}\hbar}{2m_{0}c}\rightarrow\mu_0=\textrm{magnetic moment},
\end{equation}
into (\ref{616}) one gets the Hamiltonian $H_{MR}$
\begin{equation}\label{643}
H_{MR}(p,q,\phi_{m},n;t)=\frac{p^{2}}{2m_{0}}-\mu_0\bigg((-1)^{m}b\cos(\omega t)-(-1)^{m+n}b\sin(\omega t)+(-1)^{n}B_{3}\bigg).
\end{equation}
We assume that the  Wigner function $\rho_{W}(p,q,\phi_{m},n;t)$ splits into the position-momentum part $\rho(p,q;t)$ and the spin part $\gamma(\phi_{m},n;t)$ in the following way
\begin{eqnarray}\label{644}
\rho_{W}(p,q,\phi_{m},n;t)=\rho(p,q;t)\gamma(\phi_{m},n;t),\nonumber\\
\sum_{m,n=0}^{1}\gamma(\phi_{m},n;t)=1,\hspace{1cm}\gamma^{*}=\gamma.
\end{eqnarray}
In this case the Liouville-von Neumann-Wigner equation (\ref{415}) separates as follows
\begin{subequations}\label{645}
\begin{gather}
\frac{\p\rho}{\p t}+\frac{1}{i\hbar}\left[\rho,\frac{p^{2}}{2m_{0}}\right]_{\ast}=0, \label{645a} \\
\frac{\p\gamma}{\p t}+\frac{1}{i\hbar}\left[\gamma,-\mu_0\bigg((-1)^{m}b\cos(\omega t)-(-1)^{m+n}b\sin(\omega t) +(-1)^{n}B_{3}\bigg)\right]_{\ast}=0. \label{645b}
\end{gather}
\end{subequations}
The general solution of Eq. (\ref{645a}) reads
\begin{equation}\label{646}
\rho=\rho\left(p,q-\frac{p}{m_{0}}t\right).
\end{equation}
Now we investigate the  more interesting equation (\ref{645b}). We start by introducing some new notation.
\\
Namely
\begin{eqnarray}\label{647}
\gamma_{mn}(t):=\gamma(\phi_{m},n;t),\hspace{1cm}m,n=0,1 \nonumber\\
\gamma_{0}:=2(\gamma_{00}+\gamma_{01})-1,\nonumber\\
\gamma_{1}:=2(\gamma_{00}+\gamma_{10})-1,\nonumber\\
\gamma_{2}:=2(\gamma_{00}+\gamma_{11})-1.
\end{eqnarray}
By (\ref{644}) one has
\begin{equation}\label{648}
\gamma_{00}+\gamma_{01}+\gamma_{10}+\gamma_{11}=1,
\end{equation}
and by the marginal distribution property (\ref{410c}) $\gamma_{1}(t)$ can be rewritten in the form
\begin{equation}\label{649}
\gamma_{1}(t)=2P_{+}(t)-1,
\end{equation}
where $P_{+}(t)$ is the probability that a measurement of the third component of spin at the instant $t$ gives the result $+\frac{1}{2}\hbar$. Employing (\ref{615}), (\ref{647}) and (\ref{648}), and performing simple manipulations one finds that Eq. (\ref{645b}) can be brought to the following system of equations
\begin{subequations}\label{650}
\begin{gather}
\dot{\gamma_{0}}-\frac{2\mu_0}{\hbar}(b\gamma_{1}\sin(\omega t)+B_{3}\gamma_{2})=0, \label{650a} \\
\dot{\gamma_{1}}+\frac{2\mu_0}{\hbar}b(\gamma_{0}\sin(\omega t)+\gamma_{2}\cos(\omega t))=0, \label{650b} \\
\dot{\gamma_{2}}+\frac{2\mu_0}{\hbar}(B_{3}\gamma_{0}-b\gamma_{1}\cos(\omega t))=0, \label{650c}
\end{gather}
\end{subequations}
where the overdot ``${}^\cdot$" stands for the derivative $\frac{d}{dt}$. Adding two equations, (\ref{650a}) and (\ref{650c}) multiplied by $i=\sqrt{-1}$, and denoting
\begin{equation}\label{651}
\zeta=\gamma_{0}+i\gamma_{2},
\end{equation}
we arrive at the equation
\begin{equation}\label{652}
\dot{\zeta}+i\frac{2\mu_0 B_{3}}{\hbar}\zeta-i\frac{2\mu_0 b}{\hbar}\gamma_{1}(t)\exp\{-i\omega t\}=0.
\end{equation}
The general solution of Eq. (\ref{652}) reads
\begin{equation}\label{653}
\zeta(t)=i\frac{2\mu_0 b}{\hbar}\exp\left\{-i\frac{2\mu_0 B_{3}}{\hbar}t\right\}\int dt\exp\left\{i\left(\frac{2\mu_0 B_{3}}{\hbar}-\omega\right)t\right\}\gamma_{1}(t).
\end{equation}
We are looking for the solution of Eqs. (\ref{650}) such that $P_{+}(t)$ takes the form of oscillations.
\\
We assume
\begin{equation}\label{654}
P_{+}(t)=a\sin^{2}(\Omega t), \hspace{1cm} a=\textrm{const.} \leq 1, \hspace{1cm} \Omega=\textrm{const.}
\end{equation}
Then by (\ref{649})
\begin{equation}\label{655}
\gamma_{1}(t)=2a\sin^{2}(\Omega t)-1.
\end{equation}
Inserting (\ref{655}) into (\ref{653}) one gets $\zeta(t)$ and, consequently, from (\ref{651}) also $\gamma_{0}(t)$ and $\gamma_{2}(t)$ which satisfy Eqs. (\ref{650a}) and (\ref{650c}).
However, we should also satisfy Eq. (\ref{650b}). It is easy to check that this equation can be equivalently rewritten as
\begin{equation}\label{656}
\dot{\gamma}_{1}+\frac{2\mu_0}{\hbar}b \Im\bigg\{\zeta\exp\{i\omega t\}\bigg\}=0.
\end{equation}
Finally, straightforward calculations show that the solution of system of equations (\ref{650}) (or, equivalently, of equations (\ref{652}) and (\ref{656})) under the assumption (\ref{654}) is
$$
\gamma_{0}(t)=\frac{a\hbar}{2\mu_0 b}\left\{\left(\frac{2\mu_0 B_{3}}{\hbar}-\omega\right)\cos(2\Omega t)\cos(\omega t)-2\Omega\sin(2\Omega t)\sin(\omega t)\right\}+\frac{2\mu_0 b(a-1)}{2\mu_0 B_{3} -\hbar\omega}\cos(\omega t),
$$
$$
\gamma_{1}(t)=2a\sin^{2}(\Omega t)-1,
$$
\begin{equation}\label{657}
\gamma_{2}(t)=-\frac{a\hbar}{2\mu_0 b}\left\{\left(\frac{2\mu_0 B_{3}}{\hbar}-\omega\right)\cos(2\Omega t)\sin(\omega t)+2\Omega\sin(2\Omega t)\cos(\omega t)\right\}-\frac{2\mu_0 b(a-1)}{2\mu_0 B_{3}-\hbar\omega}\sin(\omega t),
\end{equation}
with $\Omega$ given by
\begin{equation}\label{658}
\Omega=\sqrt{\left(\frac{\mu_0 b}{\hbar}\right)^{2}+\left(\frac{\mu_0 B_{3}}{\hbar}-\frac{\omega}{2}\right)^{2}}.
\end{equation}
This $\Omega$ is the celebrated \textit{Rabi frequency} \cite{Rabi,Bialynicki,Levitt,LeBellac}.
\\
From (\ref{647}) with (\ref{648}) one quickly gets the relations
\begin{equation}\label{659}
\begin{array}{ccc}
  \gamma_{00}=\frac{1}{4}(\gamma_{0}+\gamma_{1}+\gamma_{2}+1), & \hspace{1cm} & \gamma_{01}=\frac{1}{4}(\gamma_{0}-\gamma_{1}-\gamma_{2}+1), \\
  \gamma_{10}=\frac{1}{4}(\gamma_{1}-\gamma_{0}-\gamma_{2}+1), & \hspace{1cm} & \gamma_{11}=\frac{1}{4}(\gamma_{2}-\gamma_{0}-\gamma_{1}+1).
\end{array}
\end{equation}
Therefore, inserting (\ref{657}) with (\ref{658}) into (\ref{659}) we get the ``spin part" $\gamma(\phi_{m},n;t)=\gamma_{mn}(t)$, $m,n=0,1$, of the Wigner function (\ref{644}).
\\
The final step we are going to make is to find the conditions under which $\gamma(\phi_{m},n;t)=\gamma_{mn}(t)$ describes a pure spin state. Of course this is so if and only if the corresponding density operator $\widehat{\gamma}$ is the projective operator i.e. $\widehat{\gamma}^{2}=\widehat{\gamma}$.
\\
This last condition can be rewritten in terms of $\gamma(\phi_{m},n;t)$ as (see Sec. 3 and Eq. (\ref{43}) with $s+1=2$)
\begin{equation}\label{660}
\gamma * \gamma=\frac{1}{2}\gamma.
\end{equation}
Employing (\ref{613}) and (\ref{659}) after simple but rather tedious calculations one finds that Eq. (\ref{660}) is equivalent to a nice equation
\begin{equation}\label{661}
\gamma_{0}^{2}+\gamma_{1}^{2}+\gamma_{2}^{2}=1.
\end{equation}
Substituting (\ref{657}) into (\ref{661}) and using also (\ref{658}) we conclude that Eq. (\ref{661}) is in fact an algebraic equation for the amplitude $a$, and the solution reads
\begin{equation}\label{662}
a=\frac{\mu_0 b}{\hbar\Omega}.
\end{equation}
Therefore for the pure spin state, the Eq. (\ref{654}) takes the form
\begin{equation}\label{663}
P_{+}(t)=\frac{\mu_0 b}{\hbar\Omega}\sin^{2}(\Omega t),
\end{equation}
and the maximal oscillation amplitude of $P_{+}(t)$, $\frac{\mu_0 b}{\hbar\Omega}=1$, is achieved when
\begin{equation}\label{664}
\Omega=\frac{\mu_0 b}{\hbar} \Longleftrightarrow \omega=2\frac{\mu_0 B_{3}}{\hbar}.
\end{equation}
This is of course the famous \textit{magnetic resonance} effect considered in the Weyl-Wigner-Moyal formalism.
Other approach to the phase space picture of the magnetic resonance is given in \cite{Watson}.

\section{Final Remarks}

The treatment of spin within the deformation quantisation formalism is a nontrivial problem. We choose representing it in a discrete phase space.

Thus we  develop the Weyl-Wigner-Moyal formalism for the Hilbert space $L^{2}(\mathbb{R}^{3})\otimes{\mathcal{H}}^{(s+1)}$, where ${\mathcal{H}}^{(s+1)}$ is the $(s+1)$-dimensional Hilbert space, $s=1,2,\ldots$. This procedure enables us to construct a phase space picture for quantum mechanics of particle with spin $\frac{s}{2}\hbar$. The corresponding phase space of such systems is given by $\mathbb{R}^{3}\times\mathbb{R}^{3}\times \{0,...,s\} \times\{0,...,s\}$ so it consists of a continuous and a discrete part.
We obtain the expressions for the Stratonovich-Weyl quantizer, star product and Wigner functions of such systems. Formulas depend on the choice of the kernels $\mathcal{P}$ and $\mathcal{K}$ which determine the corresponding order of operators. We  use $\mathcal{P}=1$ and $\mathcal{K}$ as indicated by Eq. (\ref{54}) with which we find the expressions (\ref{57}) or (\ref{59}) for the Wigner function being the main results of this work.

The presented formalism can be employed for arbitrary values of spin. To illustrate its efficiency  we  investigate the phase space pictures of two systems.  The first one refers to the  Landau levels of spin $\frac{1}{2}$ nonrelativistic charged particle and the second one to the magnetic resonance of spin $\frac{1}{2}$ nonrelativistic uncharged particle with magnetic moment. It is shown that quite easy one can recover all Landau levels and their Wigner functions in phase space  as well as the Rabi frequency or the resonance frequency.

In the next work we are going to study the relativistic Dirac particle in terms of the Weyl-Wigner-Moyal formalism.

Finally, the results obtained throughout this work can serve as a basis for treating other physical problems in which the quantisation of discrete phase spaces could appear as in polymer quantisation.

\vskip 1.5truecm

\centerline{\bf Acknowledgments}
The work of F. J. T. was partially supported by SNI-M\'exico, COFAA-IPN and by SIP-IPN grants 20180735 and 20194924.



\end{document}